\documentclass[english,twocolumn,aps,prb,nofootinbib]{revtex4-1}
\usepackage[utf8]{inputenc}
\usepackage[utf8]{inputenc}
\usepackage[T1]{fontenc}
\usepackage{babel}
\usepackage{amsmath}
\usepackage{graphicx}
\usepackage{fancyhdr}
\usepackage{braket}
\usepackage{pagenote}
\usepackage{siunitx}
\usepackage{xcolor}
\usepackage{appendix}

\begin{document}

\title{Weyl points in systems of multiple semiconductor-superconductor quantum dots}
\author{John P. T. Stenger, David Pekker}
\affiliation{Department of Physics and Astronomy, University of Pittsburgh, Pittsburgh, PA 15260, USA}

\date{\today}

\begin{abstract}
As an analogy to the Weyl point in k-space, we search for energy levels which close at a single point as a function of a three dimensional parameter space.  Such points are topologically protected in the sense that any perturbation which acts on the two level subsystem can be corrected by tuning the control parameters.  We find that parameter controlled Weyl points are ubiquitous in semiconductor-superconductor quantum dots and that they are deeply related to Majorana zero modes.  In this paper, we present several semiconductor-superconductor quantum dot devices which host parameter controlled Weyl points.  Further, we show how these points can be observed experimentally via conductance measurements.  
\end{abstract}

\maketitle

\section{Introduction}

Like Dirac and Majorana fermions, Weyl fermions are a solution to the relativistic Dirac equation.  Furthermore, like Dirac and Majorana fermions, Weyl fermions emerge in certain solid state systems as quasiparticle modes.  In particular, they emerge in Weyl semimetals~\cite{Herring1937, Yang2011, Burkov2011, Xu2015, LV2015, Lu2015} which are characterized by a band degeneracy point in 3-dimensional k-space.  This point, known as the Weyl point, is topologically protected from environmental perturbations.  Any perturbation, which acts only on the two degenerate bands can, at most, move the Weyl point to a different location in k-space. 

Recently, it has been shown that systems of multi-terminal Josephson junctions can host Weyl points with the analogy that k-space is replaced by the space of phase differences between the terminals.~\cite{Meyer2017,Pascal2016,Xie2017,Xie2018,Yokotama2015}  Just like the traditional Weyl points, these points are immune to perturbations which, instead of removing the point, simply move it around the space of phase differences.  As charge is the conjugate variable to flux, it is natural to wonder if Weyl points can also be found by replacing k-space with charge space. 

\begin{figure}[t]
\begin{center}
\includegraphics[width=\columnwidth]{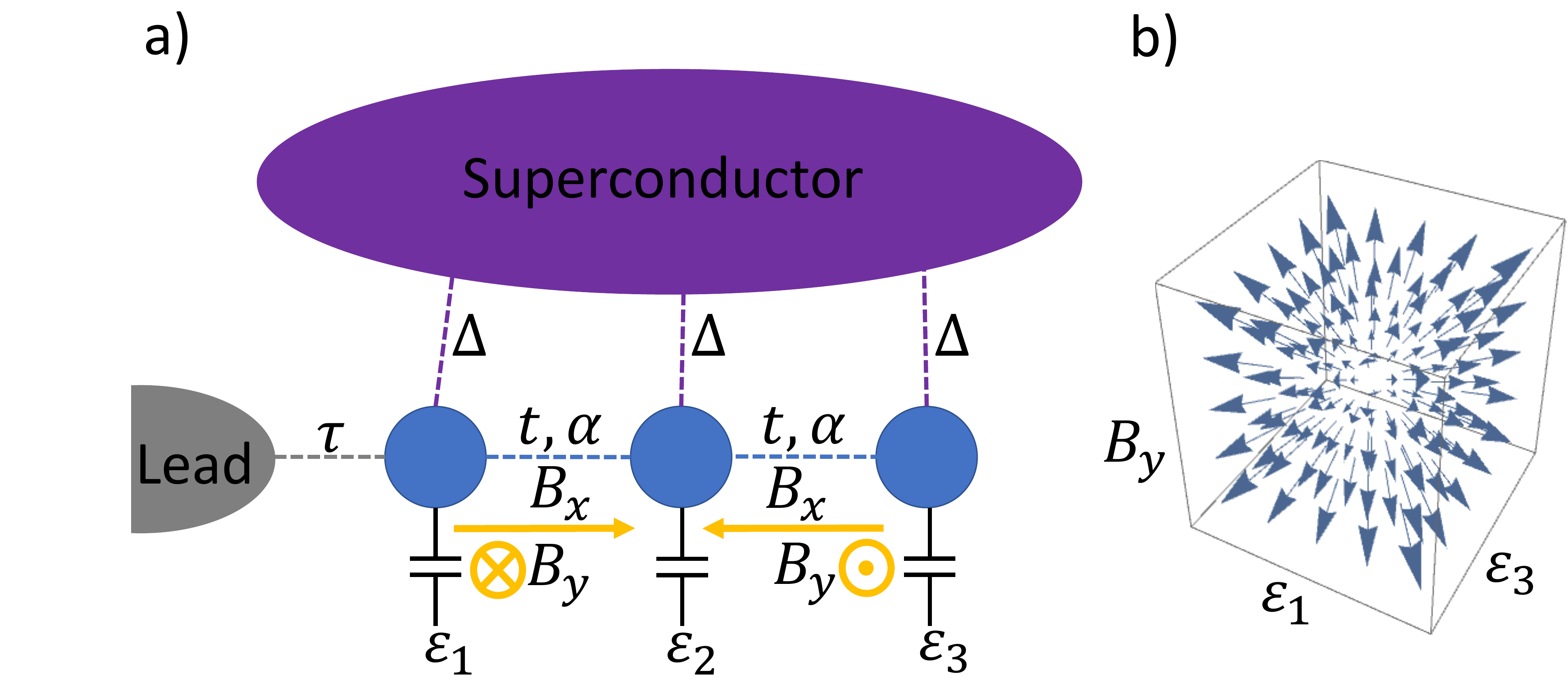}
\end{center}
\vspace{-2mm}

\caption{Schematic of a three quantum dot system.  a) three dots (blue) are proximity coupled to a parent superconductor with strength $\Delta$, the potential on each dot (labeled $\epsilon_i$) is controlled by a back gate, the dots couple to their neighbor with a hopping strength $t$ and spin orbit coupling $\alpha$, and the leftmost dot couples to a lead with strength $\tau$.  There is a magnetic field gradient in the system.  The magnetic field on the end dots points in the opposite direction and must have both an $\hat{x}$ and $\hat{y}$ component.  b) orientation of the ground state in the two level system which forms the Weyl point.  The Weyl point is controlled by the parameter space ($\epsilon_1,\epsilon_3,B_y$)}. 
\label{Fig_3}
\vspace{-3mm}
\end{figure}

In this work, we take the analogy even further by looking for Weyl points in any three dimensional space of control parameters.  In other words, we will look for Weyl Hamiltonians $H=\vec{k}\cdot\vec{\sigma}$ with $\vec{k}$ replaced by a three dimensional set of control parameters.  In particular, we will show how to search for these points in systems of three and four quantum dots like the one depicted in Fig.~\ref{Fig_3}.  The control parameters of the system can be anything that influences the Hamiltonian, however, we will attempt to use the dot potentials ($\epsilon_1,\epsilon_2,...$) when possible.  The idea being that the potentials arise from charging back gates and that charge is the conjugate variable to flux.  Recently, other control parameters have been considered such as magnetic field~\cite{Zoltan2018}.   Besides being of fundamental interest, these parameter controlled Weyl points are a signature that the chosen parameter space fully controls the Hamiltonian of a two level system.  If a Weyl point is found in any 3-dimensional space of control parameters then it is guaranteed that those parameters provide access to the entire Hilbert space for those two levels.  Because the Hilbert space is fully controllable, any unwanted perturbation to the system can be corrected. 

To find these parameter controlled Weyl points, we employ unconventional superconductivity.  Generic level crossings in quantum dot systems occur on a sheet, in a three dimensional parameter space (e.g. potentials), instead of at a single point.  However, there are special points in certain unconventional superconductor devices in which the level crossing is Weyl-like.  In particular, we will look for these Weyl points near the vicinity of separated Majorana zero modes (MZMs) in systems of quantum dots.  Kitaev predicted that MZMs will emerge from spinless p-wave superconducting chains.~\cite{Kitaev2001}  It has been shown that the Kitaev chain can be simulated by semiconductor nanowires with spin orbit coupling which are proximity coupled to a normal s-wave superconductor and which are subject to a magnetic field which is in-line with the semiconductor.~\cite{Nayak2008, Beenakker2013, Alicea2012, Sau2010, Alicea2010, Lutchyn2010, Oreg2010, Sau2010b, Ptok2017}  Signatures of the MZMs in these devices have been observed experimentally.~\cite{Mourik2012, Chen2017, Deng2016}  Another way to simulate the Kitaev chain is with chains of semiconductor-superconductor quantum dots.~\cite{Nadj2014, Fulga2013, Stenger2018, Zhang2016, Su2017}  For an infinite chain of quantum dots, the MZMs are separated in a continuous range of control parameters.  However, for a chain of a finite number of dots, as in Fig.~\ref{Fig_3}, the MZMs only separate at discrete points.  Our model is very similar to this finite Kitaev chain.  As we argue in section~\ref{WPFMSMM}, we expect to find Weyl points in the vicinity of those points where MZMs maximally separate.  There are a number of multi-dot systems in which Weyl points emerge, however, we suggest that the system in Fig.~\ref{Fig_3} is the most experimentally accessible.  

The remainder of the paper is organized as follows.  In Sec.~\ref{SM} we present two spinless models which host Weyl points, Sec.~\ref{3DC} a 3-dot chain and Sec.~\ref{4DTJ} a 4-dot tri-junction.  These models are solved analytically.  In Sec.~\ref{S3DC} we add the spin degree of freedom to the 3-dot device and discuss the similarities and differences from the spinless version.  In Sec~\ref{S3DCM} we present the model for the device, in Sec.~\ref{ELFTS3DC} we show how the Weyl point emerges and discuss which control parameters are relevant, in Sec.~\ref{MTWP} we propose an experimental method for observing the Weyl point, in Sec.~\ref{COTED} we define the charge of the Weyl point and discuss its origins, and in Sec~\ref{USOAOPDATTCP} we discuss alternative control parameters.  In Sec.~\ref{WPFMSMM} we discuss the relationship between Weyl points and Majorana operators.  In Sec.~\ref{C} we conclude.

\section{Spinless models}
\label{SM}

We begin our search with spinless, Kitaev-like models~\cite{Kitaev2001} of superconducting quantum dots.  Although, spinless models are unphysical they can be approximate solution of spinfull models under magnetic fields and are, therefore, a useful starting point. In sec.~\ref{S3DC} we will study a spinful model and show how it corresponds to the spinless version.

In Sec.~\ref{WPFMSMM} we observe that the Weyl points often accompany MZMs.  Therefore, we will narrow our search to the parameter regimes where the MZMs become maximally separated.  The concept of maximal separation is also discussed in Sec.~\ref{WPFMSMM}. 

\subsection{3-dot chain}
\label{3DC}
The first configuration is a chain of three dots for which we will try to use the chemical potential of each dot as the 3-dimensional control parameter.  However, we will show that the parameter space of potentials is not enough to find Weyl points in this device.  The Hamiltonian for this system is the following,
\begin{eqnarray}
H=\sum_{i=1}^3\epsilon_ic^{\dagger}_ic_i+t\sum_{i=1}^2(c^{\dagger}_ic_{i+1}+c^{\dagger}_{i+1}c_i)
\nonumber
\\
+\Delta\sum_{i=1}^2(e^{i\phi_i}c^{\dagger}_ic^{\dagger}_{i+1}+e^{-i\phi_i}c_{i+1}c_i)
\label{H3}
\end{eqnarray}
where $\epsilon_i$ is the potential on dot $i$, $t$ is the coupling between dots and $\Delta$ is the p-wave superconducting strength.  States with an odd number of electrons (odd parity) and an even number of electrons (even parity) are uncoupled so we can consider the two parity sectors separately.  Here we will discuss even parity.

We allow the superconducting coupling to have a different phase $\phi_i$ depending on which dots are coupled.  Consider, for example, when the two segments have opposite phase ($\phi_1=-\phi_2=-1$) and the system is at the Kitaev point $t=\Delta$.  For such a system, the even parity eigenstates at $\epsilon_1=\epsilon_2=\epsilon_3=0$ are:
\begin{eqnarray}
\ket{\psi_1}=\ket{000}+\frac{1}{\sqrt{2}}(\ket{110}-\ket{011})
\nonumber
\\
\ket{\psi_2}=\ket{101}-\frac{1}{\sqrt{2}}(\ket{110}+\ket{011})
\nonumber
\\
\ket{\psi_3}=\ket{000}-\frac{1}{\sqrt{2}}(\ket{110}-\ket{011})
\nonumber
\\
\ket{\psi_4}=\ket{101}+\frac{1}{\sqrt{2}}(\ket{110}+\ket{011})
\end{eqnarray}
where the first two states are degenerate $H\ket{\psi_{1,2}}=-\sqrt{2}\Delta\ket{\psi_{1,2}}$ and the second two states are degenerate $H\ket{\psi_{3,4}}=\sqrt{2}\Delta\ket{\psi_{3,4}}$.  For large $\Delta$, we can ignore the high energy states by projecting onto the low energy subspace.  Turning the potentials back on we find,
\begin{eqnarray}
\bra{\psi_1}H\ket{\psi_1}&=&\frac{1}{2}(\epsilon_3+\epsilon_1)+\epsilon_2-\sqrt{2}\Delta
\nonumber
\\
\bra{\psi_1}H\ket{\psi_2}&=&\frac{1}{2}(\epsilon_3-\epsilon_1)
\nonumber
\\
\bra{\psi_2}H\ket{\psi_1}&=&\frac{1}{2}(\epsilon_3-\epsilon_1)
\nonumber
\\
\bra{\psi_2}H\ket{\psi_2}&=&\frac{1}{2}(3\epsilon_3+3\epsilon_1)+\epsilon_2-\sqrt{2}\Delta
\end{eqnarray}
Representing this low energy projection in terms of the Pauli matrices, we have:
\begin{eqnarray}
H\approx \epsilon_2-\sqrt{2}\Delta+(\epsilon_3+\epsilon_1)(1-\frac{1}{2}\sigma_z)+\frac{1}{2}(\epsilon_3-\epsilon_1)\sigma_x.
\nonumber
\\
\quad
\end{eqnarray}
We see that two of the Pauli matrices can be controlled by $\epsilon_1$ and $\epsilon_3$, however, $\epsilon_2$ does not lift the degeneracy and there is no way to control $\sigma_y$ using the potentials. 

On the other hand, we can pick up $\sigma_y$ by controlling the phase of the superconductor.  Consider the perturbation Hamiltonian;
\begin{eqnarray}
H_{\delta}=\delta c^{\dagger}_1c^{\dagger}_2+\delta^{*}c_2 c_1
\end{eqnarray}
For some $\delta\ll\Delta$.  Projecting this onto the low energy subspace we get,
\begin{eqnarray}
\bra{\psi_1}H_{\delta}\ket{\psi_1}&=&\frac{1}{\sqrt{2}}(\delta+\delta^{*})
\nonumber
\\
\bra{\psi_1}H_{\delta}\ket{\psi_2}&=&-\frac{\delta^*}{\sqrt{2}}
\nonumber
\\
\bra{\psi_2}H_{\delta}\ket{\psi_1}&=&-\frac{\delta}{\sqrt{2}}
\nonumber
\\
\bra{\psi_2}H_{\delta}\ket{\psi_2}&=&0
\end{eqnarray}
In terms of the Pauli matrices, the entire Hamiltonian becomes
\begin{eqnarray}
H+H_{\delta}&\approx& \epsilon_1+\epsilon_2+\epsilon_3+\frac{1}{\sqrt{2}}\rm{Re}[\delta]-\sqrt{2}\Delta
\\
&+&\frac{1}{2}(\sqrt{2}\rm{Re}[\delta]-\epsilon_3-\epsilon_1)\sigma_z
\nonumber
\\
&+&\frac{1}{2}\left(\epsilon_3-\epsilon_1-\frac{\rm{Re}[\delta]}{\sqrt{2}}\right)\sigma_x+\frac{\rm{Im}[\delta]}{\sqrt{2}}\sigma_y
\end{eqnarray}
Therefore, we can control the full Weyl Hamiltonian by including phase control on the superconductors.

\subsection{4-dot tri-junction}
\label{4DTJ}
Given the results of the last section, it is alluring to ask if the Weyl Hamiltonian can be controlled entirely by potentials if a fourth dot is added to the system.  We will show that the answer is yes if the dot is added in a specific way.  

\begin{figure}[t]
\begin{center}
\includegraphics[width=0.7\columnwidth]{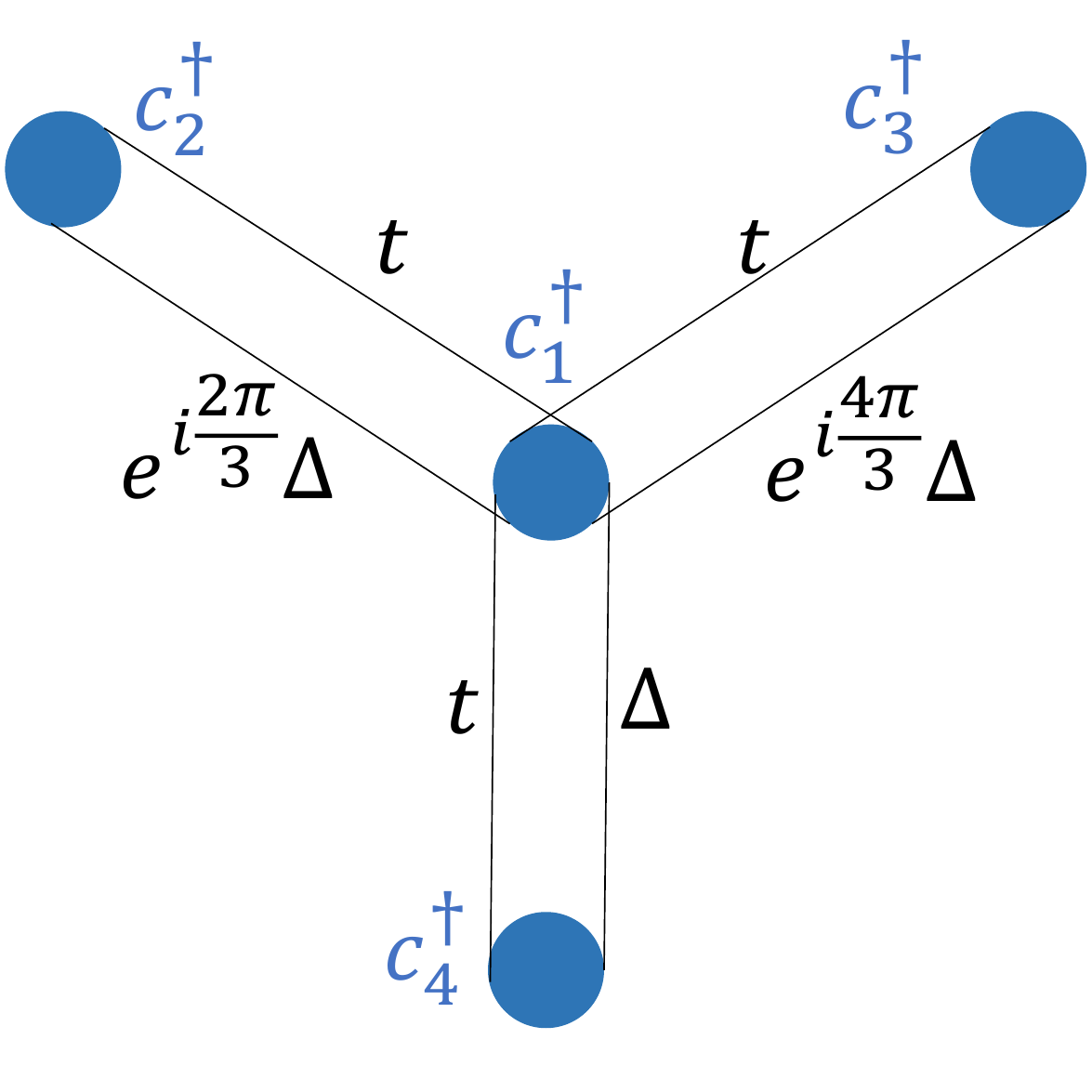}
\end{center}
\vspace{-2mm}

\caption{A four dot device that hosts Weyl points which can be fully controlled by the potentials of the outer dots.  Each outer dot is coupled to the inner dot by a hopping term of strength $t$ and a p-wave superconducting term with a particular phase and a magnitude of $\Delta$.} 
\label{Fig_2}
\vspace{-3mm}
\end{figure}

Consider the system of four dots depicted in Fig~\ref{Fig_2}.  The Hamiltonian for the system is
\begin{eqnarray}
H=\sum_{i=1}^4\epsilon_ic^{\dagger}_ic_i+t\sum_{i=2}^4(c^{\dagger}_1c_{i}+c^{\dagger}_{i}c_1)
\nonumber
\\
+\Delta\sum_{i=2}^4(e^{i\phi_i}c^{\dagger}_1c^{\dagger}_{i}+e^{-i\phi_i}c_{i}c_1)
\end{eqnarray}
where we choose $\phi_2=2\pi/3$, $\phi_3=4\pi/3$, and $\phi_4=0$.  The eigenstates at $t=\Delta$ and $\epsilon_1=\epsilon_2=\epsilon_3=\epsilon_4=0$ break up into 4 sets of 4-fold degenerate states.  Each set has 2 even parity states and 2 odd parity states.  Once again, the lowest energy states are separated from the next set of states by $\Delta$.  Therefore, we can project small perturbations onto the low energy states without worrying about the other states.  The two lowest energy, even parity states are:
\begin{eqnarray}
\ket{\psi_1}=&\frac{i}{\sqrt{6}}&\left[\sqrt{\frac{3}{2}}\ket{0000}\right.
\nonumber\\
&+&\left.\ket{1}\otimes\left(-\ket{001}+\chi^2\ket{010}+(\chi^2)^*\ket{100}\right)\right.
\nonumber\\
&+&\left.\ket{0}\otimes\left(-i\ket{110}+\chi^*\ket{101}+\chi\ket{011}\right)\frac{1}{\sqrt{2}}\right]
\nonumber\\
\ket{\psi_2}=&\frac{1}{\sqrt{6}}&\left[\sqrt{\frac{3}{2}}\ket{1111}\right.
\nonumber\\
&+&\left.\ket{0}\otimes\left(-\ket{110}-(\chi^2)^*\ket{101}+\chi^2\ket{011}\right)\right.
\nonumber\\
&+&\left.\ket{1}\otimes\left(-i\ket{001}+\chi\ket{010}-\chi^*\ket{100}\right)\frac{1}{\sqrt{2}}\right].
\nonumber\\
\quad
\end{eqnarray}
where $\chi=\exp(i\pi/6)$.  Using this basis, we can write the low energy projection of the Hamiltonian in terms of the Pauli matrices.  Keeping $\epsilon_1=0$ but turning on the other three potentials, we have: 
\begin{eqnarray}
H\approx&-&\frac{1}{6}(\epsilon_2+\epsilon_3+\epsilon_4)\sigma_z
\nonumber
\\
&-&\frac{1}{6\sqrt{2}}\left(\epsilon_2+\epsilon_3-2\epsilon_4\right)\sigma_x
\nonumber
\\
&-&\frac{1}{2\sqrt{6}}\left(\epsilon_2-\epsilon_3\right)\sigma_y
\end{eqnarray}
where we have dropped the terms which multiply the identity matrix.  We see that each of the three Pauli matrices can be controlled by the potentials under the three outer dots.  Since the Hilbert space of two level systems is exhausted by the Pauli matrices, any small perturbation to the system can be corrected by tuning these three potentials.  In other words, the Weyl point cannot be removed form the three dimensional space of potentials by perturbing the system with energies less than $\Delta$.  However, if any two superconducting phases are the same then we lose control over one of the Pauli matrices. 

\section{Spinful 3-dot chain}
\label{S3DC}

Kitaev-like spinless models can be obtained as the low energy limit of a spinfull model in the presence of spin orbit coupling and a magnetic field.  In the last section, we saw that in order to control all three Pauli matrices in the spinless 3-dot chain we needed to control the superconducting phase.  However, we will see that instead of controlling the phase difference of superconductors, we can instead control the orientation of either the spin orbit coupling or the magnetic field.  We suggest that the spinful 3-dot device (see Fig.~\ref{Fig_3}) is the most easily accessible system for experimental study and that the magnetic field orientation is the most easily accessible control parameter.

\subsection{Spinful 3-dot chain model}
\label{S3DCM}
We study a three dot Hamiltonian with on site potential, nearest neighbor hopping, spin orbit coupling, proximity induced Andreev reflection, a magnetic field, and interactions (see Fig.~\ref{Fig_3}).  This is the most general model which hosts MZMs.  The interactions are not necessary to observe a Weyl point but we include it to demonstrate the stability of the Weyl point. 
\begin{eqnarray}
H=H_{\epsilon}+H_{\Delta}+H_{t}+H_{\alpha}+H_{B}+H_{U}
\end{eqnarray}
Here the onsite potential Hamiltonian is,
\begin{eqnarray}
H_{\epsilon}=\sum_{i,\sigma}\epsilon_i c^{\dagger}_{i\sigma}c_{i\sigma}
\end{eqnarray}
where $\epsilon_i$ is the onsite potential, $i\in\{1,2,3\}$ runs over the three dots, and $\sigma\in\{\uparrow,\downarrow\}$.  The Andreev Hamiltonian is,
\begin{eqnarray}
H_{\Delta}=\sum_{i}\left(\Delta_i c^{\dagger}_{i,\uparrow}c^{\dagger}_{i,\downarrow}+\Delta_i^{*}c_{i,\downarrow}c_{,i\uparrow}\right)
\end{eqnarray}
where $\Delta_i$ is the induced Andreev reflection amplitude on dot $i$.  

The hopping Hamiltonian is,
\begin{eqnarray}
H_{t}=\sum_{i,\sigma}t_{i}\left(c^{\dagger}_{i,\sigma}c_{i+1,\sigma}+c^{\dagger}_{i+1,\sigma}c_{i\sigma}\right)
\end{eqnarray}
where $t_i$ is the nearest neighbor hopping strength between dot $i$ and dot $i+1$.  The spin-orbit coupling term is broken up into its two component directions:
\begin{eqnarray}
H_{\alpha}=H_{\alpha_x}+H_{\alpha_y}
\end{eqnarray}
with
\begin{eqnarray}
H_{\alpha_x}=\sum_i \alpha_i\cos(\xi_i)\left(c^{\dagger}_{i,\uparrow}c_{i+1,\downarrow}-c^{\dagger}_{i,\downarrow}c_{i+1,\uparrow}+h.c.\right)
\nonumber
\\
H_{\alpha_y}=i\sum_i \alpha_i\sin(\xi_i)\left(c^{\dagger}_{i,\uparrow}c_{i+1,\downarrow}+c^{\dagger}_{i,\downarrow}c_{i+1,\uparrow}+h.c.\right)
\nonumber
\\
\quad
\label{Haxy}
\end{eqnarray}
where $\alpha_i$ is the overall strength of the spin orbit coupling between dot $i$ and $i+1$, and $\xi_i$ is the angle that a line between the two dots makes with the x-axis.  The magnetic field Hamiltonian is also broken up into its component directions,
\begin{eqnarray}
H_{B}=H_{B_x}+H_{B_y}+H_{B_z}
\end{eqnarray}
with
\begin{eqnarray}
&H_{B_x}&=\sum_i B_i\sin(\theta_i)\cos(\phi_i)\left(c^{\dagger}_{i,\uparrow}c_{i,\downarrow}+c^{\dagger}_{i,\downarrow}c_{i,\uparrow}\right)
\nonumber
\\
&H_{B_y}&=i\sum_i B_i\sin(\theta_i)\sin(\phi_i)\left(c^{\dagger}_{i,\uparrow}c_{i,\downarrow}-c^{\dagger}_{i,\downarrow}c_{i,\uparrow}\right)
\nonumber
\\
&H_{B_z}&=\sum_i B_i\cos(\theta_i)\left(c^{\dagger}_{i,\uparrow}c_{i,\uparrow}+c^{\dagger}_{i,\downarrow}c_{i,\downarrow}\right)
\end{eqnarray}
where $B_i$ is the magnitude of the magnetic field on dot $i$, $\theta_i$ is the polar angle, and $\phi_i$ is the azimuthal angle of the field on dot $i$.  The interaction term is,
\begin{eqnarray}
H_U=\sum_iU_ic^{\dagger}_{i,\uparrow}c_{i,\uparrow}c^{\dagger}_{i,\downarrow}c_{i,\downarrow}
\end{eqnarray}
where $U_i$ is the interaction strength on dot $i$.

In what follows, we will assume that all three dots are isomorphic except where otherwise specified and we will refer to isomorphic parameters by dropping the site index (e.g. we will refer to $\Delta$ when $\Delta_1=\Delta_2=\Delta_3$).  
\begin{figure}[t]
\begin{center}
\includegraphics[width=\columnwidth]{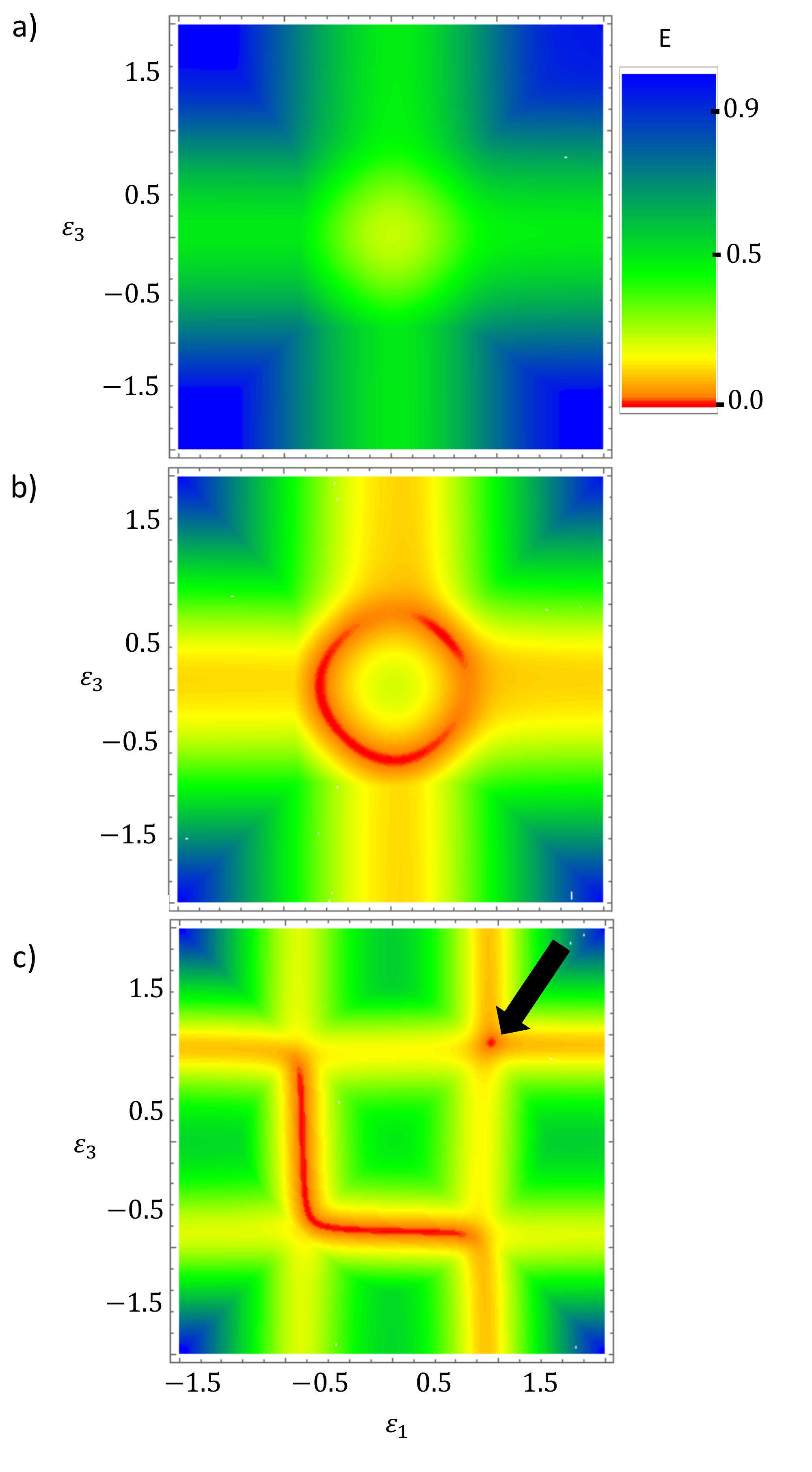}
\end{center}
\vspace{-2mm}

\caption{Emergence of the Weyl point.  Each panel shows the energy difference, E, between the first two even parity states as a function of the potential under dot 1, $\epsilon_1$, and dot 3, $\epsilon_3$, for different values of the magnetic field, $B_x$.  All energy parameters are in units of the induced gap, $2\Delta$.  (a) $B_x=0.6 \Delta$. (b) $B_x=1.4 \Delta$. (c) $B_x=2\Delta$.  For magnetic fields below the Andreev coupling $B_x<\Delta$, as in panel (a), the energy levels are gaped in the $\epsilon_1$-$\epsilon_3$ plane and there is no Weyl point.  The Weyl point emerges above $B_X>\Delta$, however, near $B_x=\Delta$, as in panel (b), the point is smeared.  On the other hand, the Weyl point is sharp in the top right corner of panel (c).} 
\label{Fig_4}
\vspace{-3mm}
\end{figure}

\begin{figure}[t]
\begin{center}
\includegraphics[width=\columnwidth]{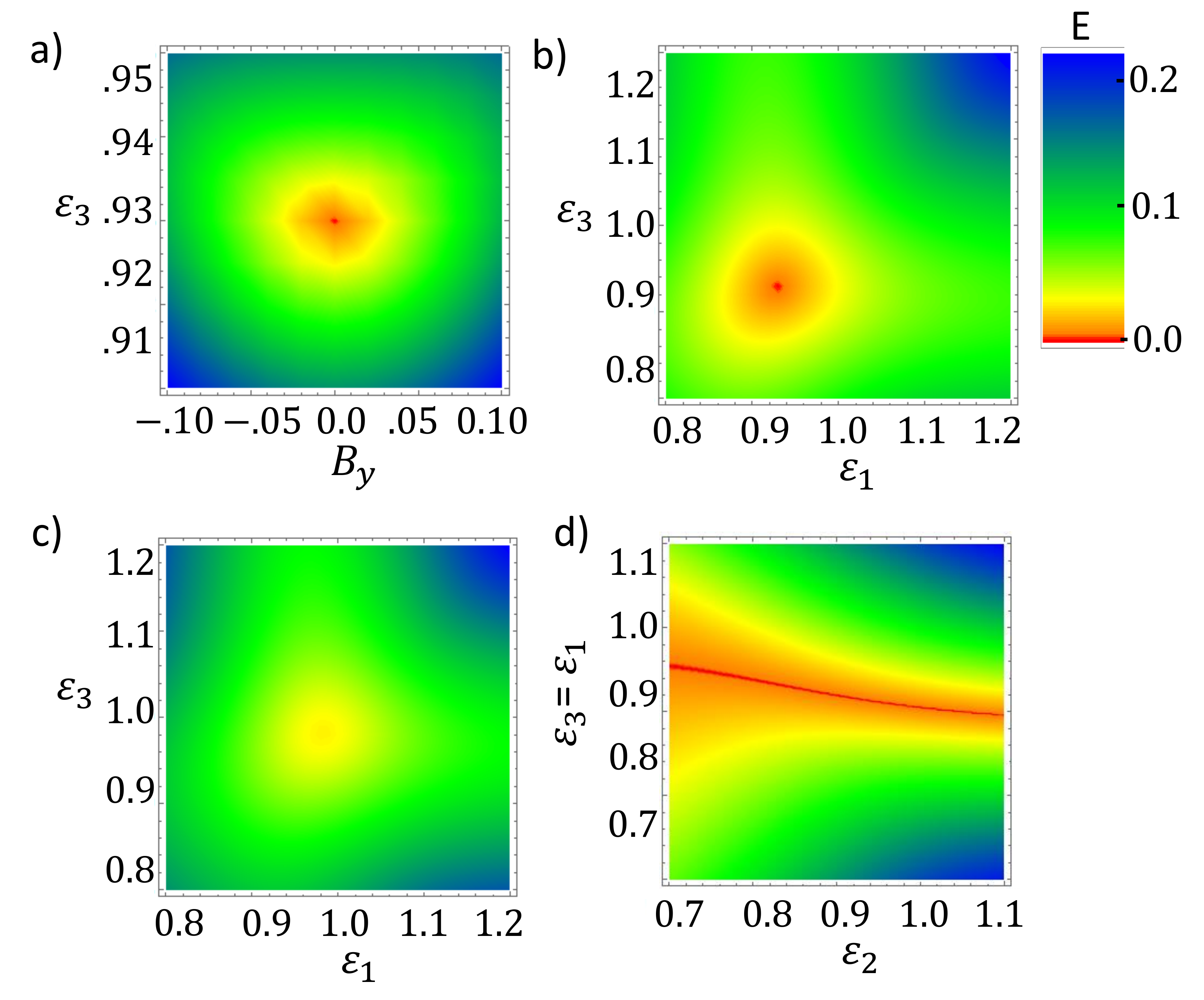}
\end{center}
\vspace{-2mm}

\caption{Mapping out the Weyl point.  Each panel shows the energy difference, E, between the first two even parity states as a function of various control parameters which are defined in the main text.  All parameters are in units of $2\Delta$.   (a) $\epsilon_1=1.86\Delta$, (b) $B_y=0$, (c) $B_y=0.6\Delta$, (d) $B_y=0$.  Panels (a) and (b) show two dimensional slices of the Weyl cone.  These show that all three parameters $(\epsilon_1, \epsilon_2, B_y)$ control the energy level separation.  However, to prove that there is a Weyl point we need panel (c) which shows that $B_y$ opens the gap everywhere in the 2D $(\epsilon_1, \epsilon_2)$ space.  Panel (d) shows that $\epsilon_2$ does not open the gap and therefore cannot be used as a control parameter.} 
\label{Fig_5}
\vspace{-3mm}
\end{figure}

\subsection{Energy levels for the spinfull 3-dot chain}
\label{ELFTS3DC}

\begin{figure}[t]
\begin{center}
\includegraphics[width=\columnwidth]{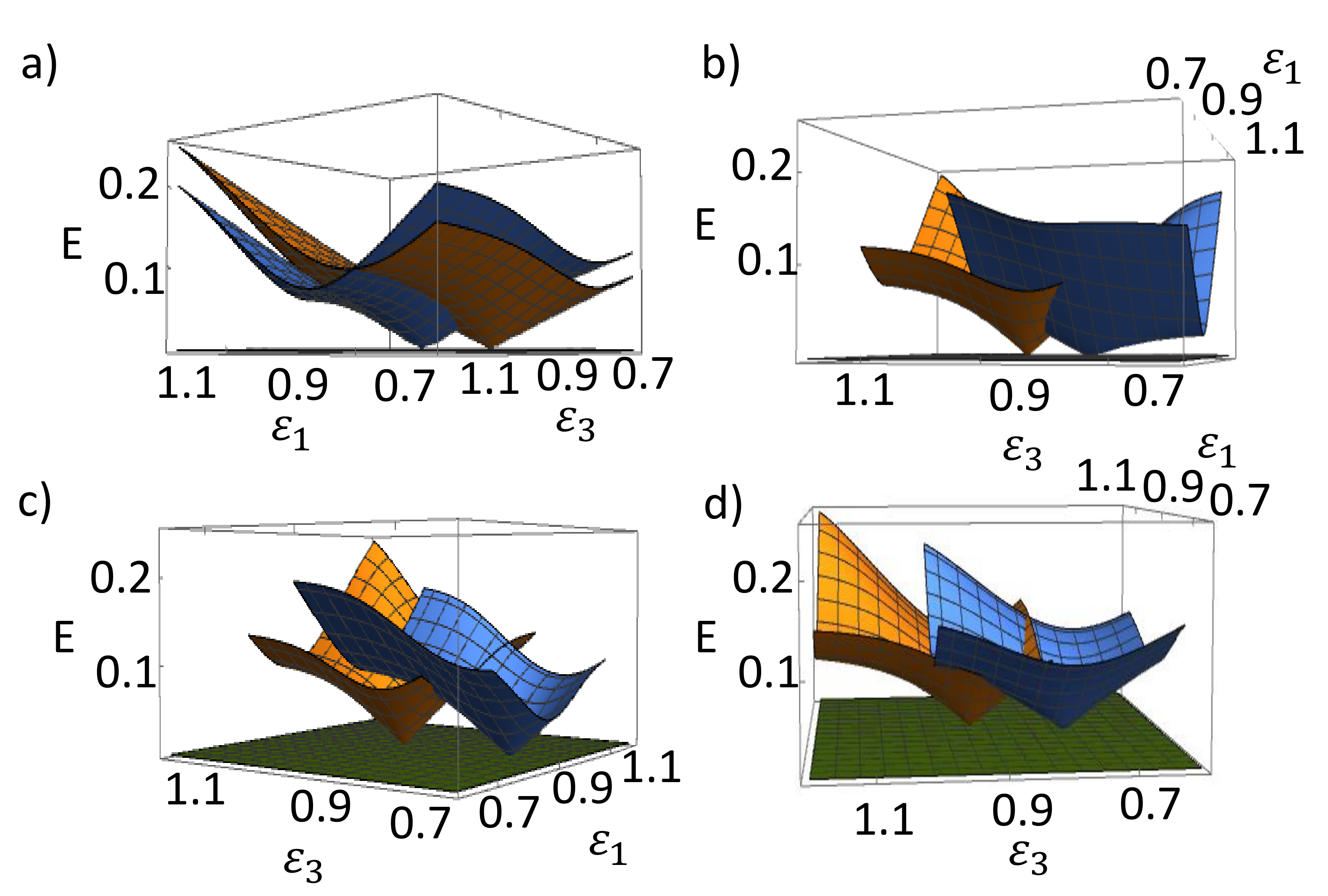}
\end{center}
\vspace{-2mm}

\caption{Protection of the Weyl point.  Each panel shows the energy difference, E, between the first two even parity states under various small perturbations overlaid with the unperturbed case.  All parameters are in units of $2\Delta$.  The yellow curve in each panel is the unperturbed case.  The blue curve is for (a) $U=\Delta$, (b) $t_1=0.5t_2=0.5t_3$ and $\alpha_1=0.5\alpha_2=0.5\alpha_3$, (c) $B_3=0.8B_2=0.8B_3$, (d) $\Delta_2=1.2\Delta_1$ and $\Delta_3=1.4\Delta_1$} 
\label{Fig_6}
\vspace{-3mm}
\end{figure}

In order to obtain Weyl points, we want to mimic the spinless model already discussed (Eq.~\ref{H3}).  There we had that the two p-wave pairing terms had opposite sign.  We can mimic this behaviour, without applying a phase difference directly to the superconductors, by instead controlling the magnetic field orientation under each dot.  We take this approach as we expect it to be experimentally less challenging than controlling the phase differences.  In subsection~\ref{USOAOPDATTCP} we show that controlling the superconducting phase and the spin-orbit angle are both viable alternatives to the magnetic field orientation.

Let us take $\theta=0$ and $\phi_1=\pi$ while $\phi_2=\phi_3=0$.  Figure~\ref{Fig_4} shows the energy difference between the first two even parity states as a function of $\epsilon_1$ and $\epsilon_3$.  Just like in the spinless case, $\epsilon_2$ does not open or close the Weyl point.  We see that the Weyl point drops down and persists for $B>\Delta$.  This topological phase transition coincides with the appearance of maximally separated MZMs.  

As we have mentioned, we want to use a magnetic field gradient as the third control parameter.  We have already used a magnetic field gradient in the $\hat{x}$-direction to drive the topological phase transition.  Therefore we use a gradient in the $\hat{y}$-direction as the third control parameter.   Keeping $\theta=0$, let us take $\phi_1=\Phi+\pi$, $\phi_2=0$, and $\phi_3=-\Phi$ and let us define $B_x=B\cos(\Phi)$ and $B_y=B\sin(\Phi)$.  Keeping $B_x$ constant, we can control the third axis of the Weyl point with $B_y$.  

Although the magnetic field has to rotate over a rather short distance, the gradient should be experimentally achievable.  The strength of the gradient depends on the magnetic g-factor and the distance between the dots.  Although there is no strict requirements for the distance between the dots, let us say that the dots are about $1~\mu$m apart which is the a typical length for Majorana nanowires.  Then from experimentally observed g-factors~\cite{Chen2017}, we need a gradient in the x-direction of about 0.5 T/$\mu$m.  This is rather large but is predicted to be achievable using nanomagnets~\cite{Maurer2018}.  On the other hand, the magnetic gradient in the y-direction can be much smaller and should be achievable using electromagnets~\cite{Drndic1998}.  

Figure~\ref{Fig_5} shows the energy difference between the first two even parity states as a function of various control parameters. Panel (a) shows the Weyl cone as a function of $B_y$ and $\epsilon_3$ while panel (b) shows the cone as a function of $\epsilon_1$ and $\epsilon_3$.  We conclude that we have linear dispersion as a function of all three control parameters ($\epsilon_1,\epsilon_3,B_y$).  Panel (c) shows that the Weyl point vanishes from the two dimensional $\epsilon_1$-$\epsilon_3$ space when $B_y$ is turned on.  In other words, there is no line $B_y=a\epsilon_{1}+b\epsilon_2$ (for arbitrary $a$ and $b$) on which the point stays closed.  Panel (d) shows that $\epsilon_2$ does not open or close the Weyl point.  The Weyl point's immunity to $\epsilon_2$ means that this potential cannot be used as a third control parameter.  

The immunity of the Weyl point to $\epsilon_2$ is also an example of the topological protection of the point.  Just as the standard Weyl point is a source of Berry curvature in k-space, the parameter controlled Weyl point is a source of the curvature in parameter space.  Since we have control over all three Pauli matrices, there are no small perturbations (compared to $\Delta$) which removes the Weyl point from the 3-dimensional parameter space.  On the other hand, perturbations larger than $\Delta$ can be damaging in two ways.  First, they could simply move the Weyl point to locations in parameter space which are out of the range of realistic tuning parameters.  Second, they can close the gap which could potentially destroy they Weyl point.  Figure~\ref{Fig_6} shows the effect of several types of perturbations.  In each panel, the yellow curve is the unperturbed energy difference between the first and second even parity state and the blue curve is the perturbed energy difference.  In fact, we see that even some perturbations on the order of $\Delta$ do not remove the Weyl point.  In Fig.~\ref{Fig_6}a, for example, the interaction strength is tuned to $U=\Delta$.  In Fig.~\ref{Fig_6}b we perturb the relative hopping strength between dots.  This type of perturbation, which breaks the isomorphism between dots, takes the Weyl point off of the diagonal $\epsilon_1=\epsilon_3$.  This behaviour can also be seen in Fig~\ref{Fig_6}c and Fig.~\ref{Fig_6}d where we break the magnetic and superconducting isomorphisms respectively.

\subsection{Measuring the Weyl point}
\label{MTWP}

The Weyl point in the three dot chain can be observed experimentally by tunnel coupling a metallic lead to one of the dots in the system.  In this setup, electrons tunnel into the dots from the lead and then Andreev reflect off of the superconductor contacts.  We calculate the differential conductance using the well established master equation formalism.~\cite{Su2017}
\begin{eqnarray}
\dot{P_i}=\sum_{j}\left(\Gamma^{j\rightarrow i}P_j-\Gamma^{i \rightarrow j}P_{i}\right)
\end{eqnarray}
where $P_i$ is the probability of the system being in state $i$.  We use the steady state approximation $\dot{P_i}=0$ and connect the lead to the first dot.  The rates $\Gamma^{i\rightarrow j}$ are given by,
\begin{eqnarray}
\Gamma^{i\rightarrow j}=\Gamma^{i\rightarrow j}_p+\Gamma^{i \rightarrow j}_h
\end{eqnarray}
with
\begin{eqnarray}
&\Gamma^{i\rightarrow j}_p&=\tau^2f(E_j-E_i-eV)\sum_{\sigma}|\bra{j}c^{\dagger}_{1\sigma}\ket{i}|^2
\\
&\Gamma^{i\rightarrow j}_h&=\tau^2(1-f(E_j-E_i-eV))\sum_{\sigma}|\bra{j}c_{1\sigma}\ket{i}|^2
\nonumber
\end{eqnarray}
where $\ket{i}$ is the eigenvector with eigenvalue $E_i$, $f(\omega)$ is the Fermi distribution, and $\tau$ is the coupling between the lead and the first dot.  Once we know all of the probabilities $P_i$, we can calculate the current using,
\begin{eqnarray}
I=\sum_{i,j}P_i\left(\Gamma^{i\rightarrow j}_p-\Gamma^{i\rightarrow j}_h\right)
\end{eqnarray}
and the differential conductance is simply $dI/dV$ which we calculate numerically.

\begin{figure}[t]
\begin{center}
\includegraphics[width=\columnwidth]{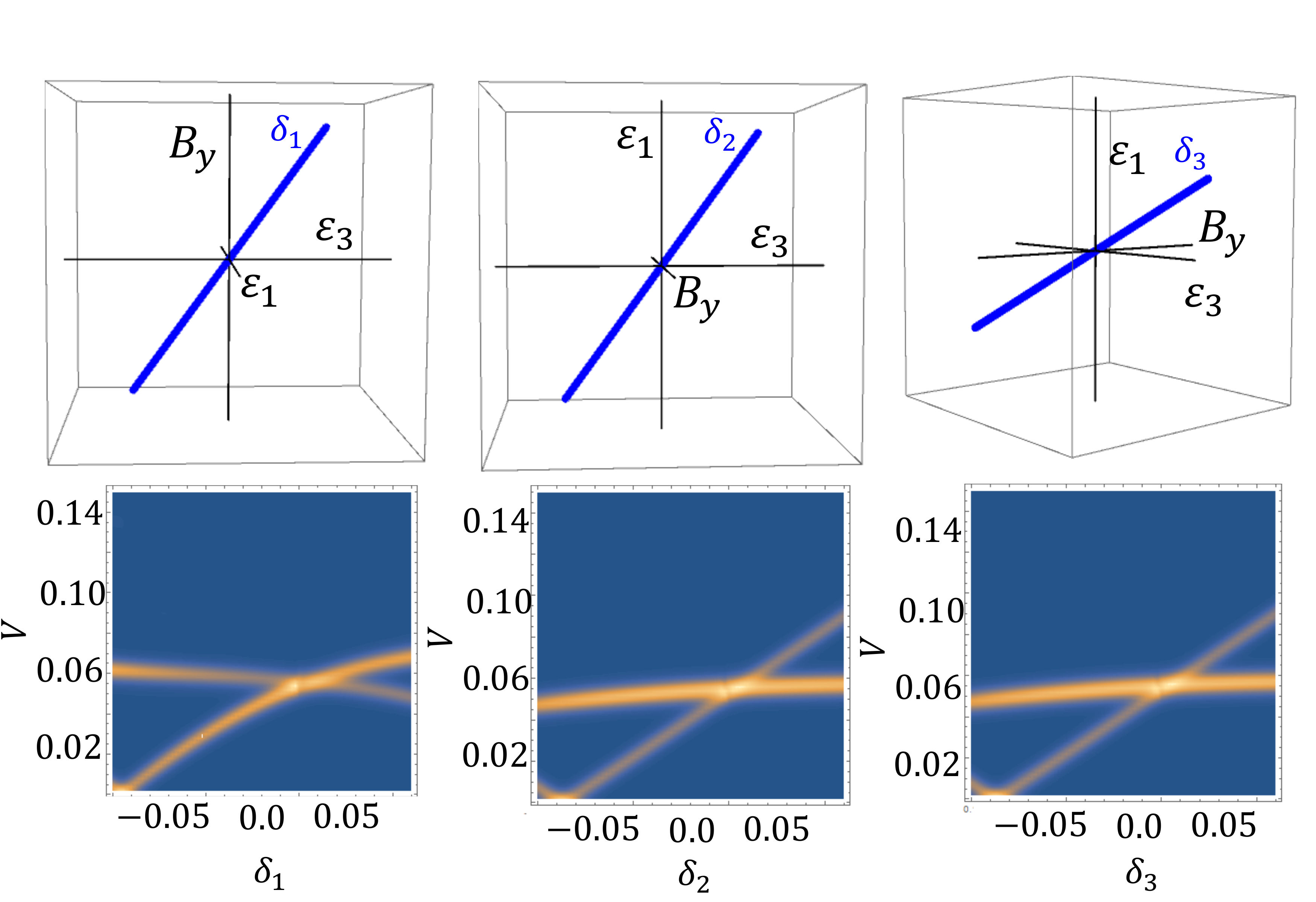}
\end{center}
\vspace{-2mm}

\caption{Measuring the Weyl point via tunnel conductance. Each scan is done in a different direction in the three dimensional parameter space.  The top row shows the different paths $\delta_i(\epsilon_1,\epsilon_2,B_y)$ in the 3-dimensional parameter space that are used.  The bottom row shows the differential conductance as a function of bias voltage $V$ and the corresponding parameter $\delta_i$.  All parameters are in units of 2$\Delta$.} 
\label{Fig_7}
\vspace{-3mm}
\end{figure}

Results of the transport calculation are shown in Fig.~\ref{Fig_7}.  The goal in experiment will be to check that the energy gap, between the first and second even parity states, opens along all paths through the center of the three dimensional parameter space.  The data in Fig~\ref{Fig_7} is taken along the paths $\delta_i(\epsilon_1,\epsilon_2,B_y)$ depicted on the top row.  It happens that the ground state of this system is an odd parity state which is separated in energy from the first even parity state at the Weyl point.  Therefore, we see two finite bias conductance  curves (bottom row) which cross at the Weyl point and reopen regardless of the parameter path  that is chosen.   

Note that the MZMs maximally separate when the odd parity ground state and one of the even parity states are degenerate.  This can be seen in Fig~\ref{Fig_7} at around $\delta\approx-0.1$ in all three panels.  As expected (see Sec.~\ref{WPFMSMM}), the Weyl point appears in the vicinity of maximally separated MZMs.       

\subsection{Curvature of the energy degeneracies}
\label{COTED}
Weyl points in k-space are characterized by a non-zero, integer Chern number. Similarly, we can define an analogous integer for the parameter controlled Weyl point by integrating over parameter space instead of k-space.  Just like the normal Chern number, we can define this integral using a gauge field given by the Berry connection.
\begin{equation}
    \vec{A}_{n}(\vec{r})=i\bra{n(\vec{r})}\vec{\nabla}_r\ket{n(\vec{r})}
\end{equation}
where $\ket{n}$ is the $n^{\rm{th}}$ energy level eigenstate and $\vec{r}=\{\epsilon_1,\epsilon_3,B_y\}$.  The local curvature is then given by the curl of this gauge field,
\begin{equation}
    \vec{\Omega}_n(\vec{r})=\vec{\nabla}_r\times\vec{A}_n(\vec{r}).
\end{equation}
Integrating the curvature on a closed surface, we get an integer which is zero if the surface does not enclose a degeneracies.  We will call this integer the charge of a degeneracy surface or point,  
\begin{equation}
    \mathcal{C}_n=\frac{1}{2\pi}\oint d\vec{S}\cdot\vec{\Omega}_n.
\end{equation}
Let us apply this formalism to better understand the parameter controlled Weyl point of our three dot device (seen in e.g. Fig~\ref{Fig_4}c).  

\begin{figure}[t]
\begin{center}
\includegraphics[width=\columnwidth]{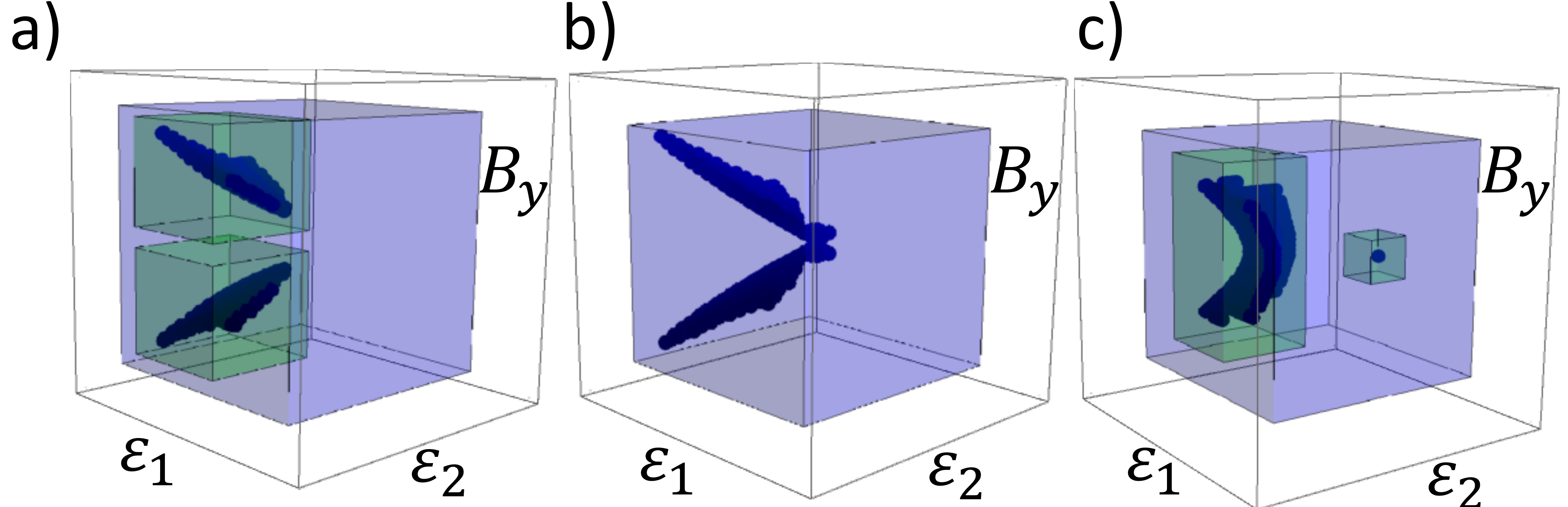}
\end{center}
\vspace{-2mm}

\caption{Surfaces in the three dimensional parameter space where the first and second even parity states are degenerate.  Going from right to left $B_x=0.6\Delta,~1.4\Delta,~2.0\Delta$, matching the values in Fig.~\ref{Fig_4}.  The global charge, calculated by integrating over the blue box, is always $\mathcal{C}_0=+2$.  The charges enclosed by the green boxes are $\mathcal{C}_0=+1$ for both panels (a) and (c).}
\label{Fig_11}
\vspace{-3mm}
\end{figure}

In Fig.~\ref{Fig_11} we show the surfaces of degeneracy between the first two even parity states.  We see that the Weyl point emerges when two positively charged degeneracy surfaces collide in the $B_y=0.0$ plane.  After the collision, the Weyl point breaks off from the surface and takes half of the total charge with it.  The global charge is unchanged throughout the entire process.  In principle, the positively charged Weyl point could be removed by combining with a negatively charged degeneracy point.  However, there are no negative charges within the parameter regimes we have scanned.  

\subsection{Using spin-orbit angle or phase differences as the third control parameter}
\label{USOAOPDATTCP}
So far we have used the magnetic field gradient as the third control parameter.  However, there are at least two other possible choices, namely a superconducting phase difference between dots or a spin-orbit angle.  These parameters are likely more difficult to control experimentally than the magnetic field gradient but they are equally valid.  

\begin{figure}[t]
\begin{center}
\includegraphics[width=\columnwidth]{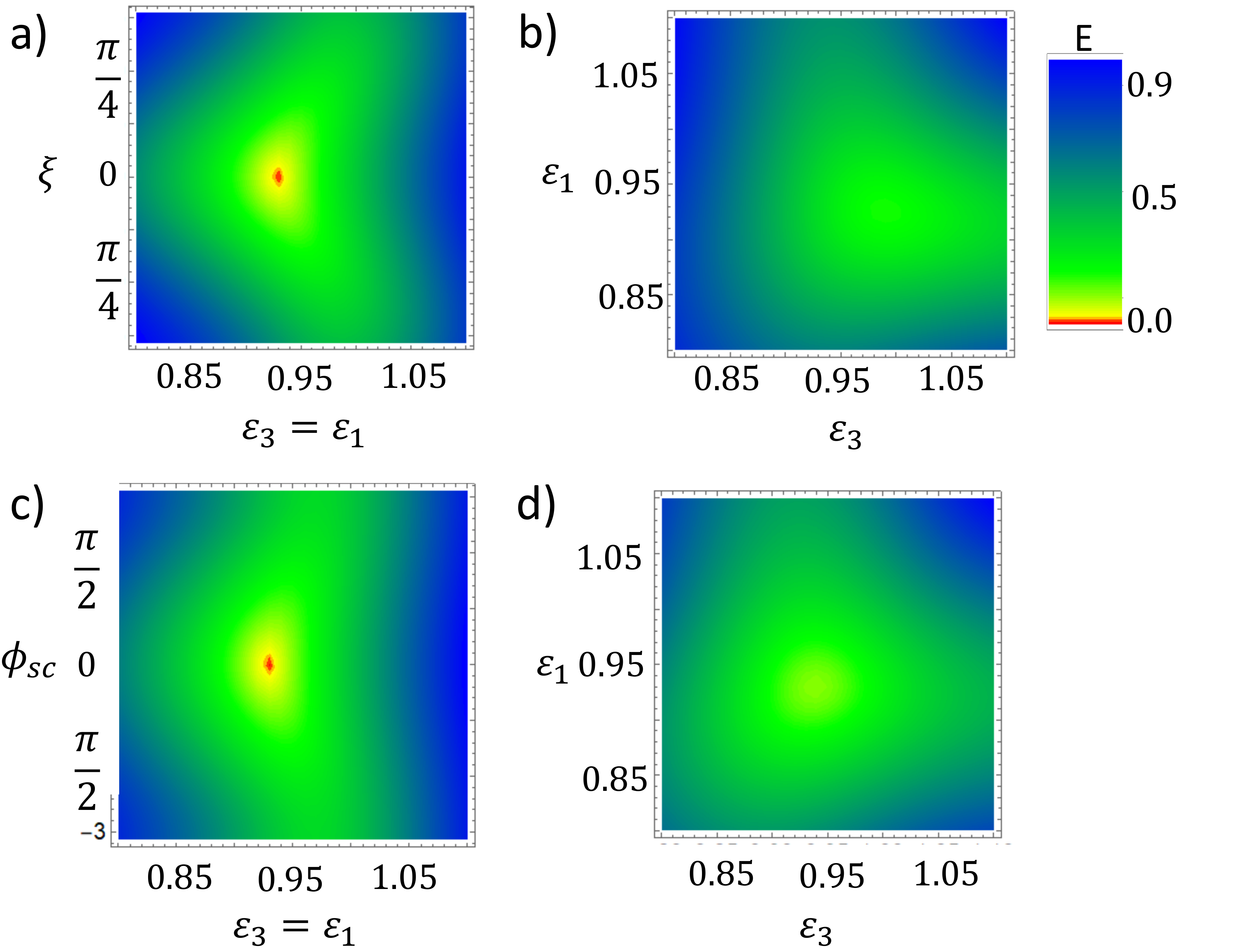}
\end{center}
\vspace{-2mm}

\caption{Energy difference between the first two even parity states. a) as a function of the spin-orbit angle $\xi_2$ and the potentials $\epsilon_3=\epsilon_1$.  b) as a function of the dot potentials $\epsilon_1$ and $\epsilon_2$ for $\xi=\pi/4$.  c) as a function of the phase of the third superconducting contact $\phi_{sc}=-i\ln(\Delta_3/|\Delta_3|)$ while the other phases are zero, $\rm{Im}(\Delta_1)=\rm{Im}(\Delta_2)=0$.  d) as a function of the dot potentials for $\phi_{\rm sc}=\pi/4$}
\label{Fig_8}
\vspace{-3mm}
\end{figure}

In Fig.~\ref{Fig_8} we plot the energy difference between the first two even parity states showing that both the superconducting phase and the spin-orbit angle can be used to control the Weyl point.  Figure~\ref{Fig_8}a shows the Weyl point in the parameter space $(\epsilon_1,\epsilon_3,\xi_2)$ where $\xi_2$ is the spin-orbit angle from Eq.~\ref{Haxy}.  We see in Fig.~\ref{Fig_8}b that $\xi_2$ opens the gap everywhere in the 2-dimensional $(\epsilon_1,\epsilon_3)$ space meaning that it can truly be used as the third control parameter.  Similarly, in Fig.~\ref{Fig_8}c we see the Weyl point in the parameter space $(\epsilon_1, \epsilon_3, \phi_{\rm sc})$ where $\phi_{\rm sc}=-i\ln(\Delta_1/|\Delta_1|)$ is the superconducting phase angle on the first dot.  We set the phase on the other dots to zero.  In Fig.~\ref{Fig_8}d we see that $\phi_{\rm sc}$ also opens the gap everywhere in the 2-dimensional space $(\epsilon_1,\epsilon_3)$ and is therefore another valid control parameter.  

\section{Weyl point from maximally separated Majorana modes}
\label{WPFMSMM}

We have seen that Weyl points emerge in several p-wave superconducting devices.  This is not an accident.  In fact, Weyl Hamiltonians arise naturally from a comparison of the algebra of Majorana operators and that of the Pauli matrices (i.e. the quaternion algebra).  Indeed, even the Majorana operators themselves form a quaternion algebra.  Consider a pair of Majorana operators $\gamma_x$ and $\gamma_y$ which form the parity operator $P=i\gamma_x\gamma_y$.  Any pair of these three operators $\gamma_x$, $\gamma_y$, and $P$ multiplied together gives the third in exactly the same way as the Pauli matrices.  Therefore, the Pauli matrices are a representation of a pair of Majorana operators and their parity.  However, Hamiltonians come with pairs of Majorana operators, not single operators.  Therefore, to reproduce the Weyl Hamiltonian, we need to find a quaternion algebra that involves only even numbers of Majorana operators. 

\begin{figure}[t]
\begin{center}
\includegraphics[width=\columnwidth]{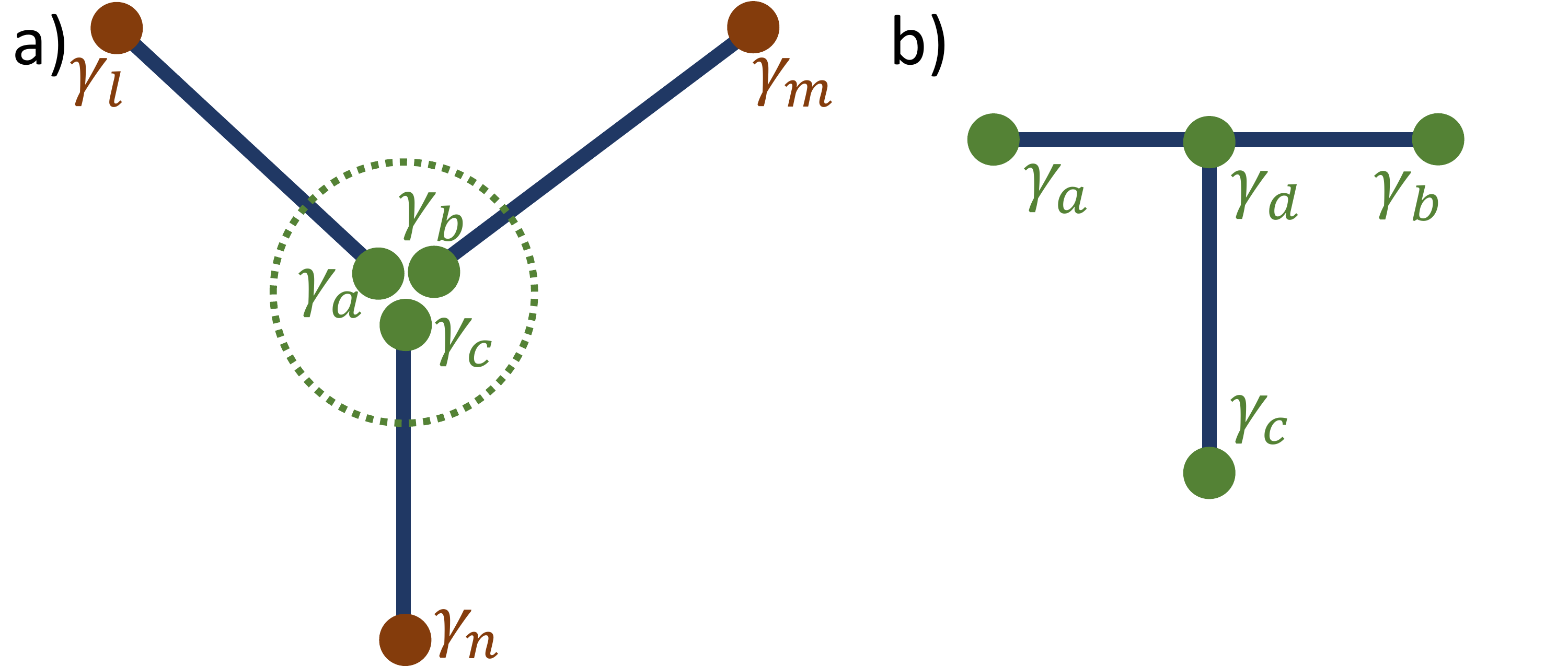}
\end{center}
\vspace{-2mm}

\caption{Topological superconducting nanowires (blue) with Majorana end modes (red and green) in configurations that host Weyl points.  The green Majoranas are active in the formation of the Weyl point while the red Majoranas are auxiliary.} 
\label{Fig_1}
\vspace{-3mm}
\end{figure}

One such algebra involves pairs of three different Majorana operators.  Consider for example the system in Fig.~\ref{Fig_1}a where there are six Majorana operators but only three of them are coupled ($\gamma_a$, $\gamma_b$, $\gamma_c$).  Now consider the operators $i\gamma_a\gamma_b$, $i\gamma_a\gamma_c$, and $i\gamma_b\gamma_c$.  Once again, these operators form a quaternion algebra and can be represented by the Pauli matrices.  Therefore, the Hamiltonian
\begin{eqnarray}
\label{Ha}
H_a=it_{ab}\gamma_a\gamma_b+it_{ac}\gamma_a\gamma_c+it_{bc}\gamma_b\gamma_c 
\end{eqnarray}
can be represented by the Pauli matrices,
\begin{eqnarray}
\bar{H}_a=\vec{t}\cdot\vec{\sigma}
\end{eqnarray}
where $\vec{t}=(t_{ab},t_{ac},t_{bc})$.  Of course, the Hilbert space of $H_a$ is twice that of $\bar{H}_a$.  However, the even and odd subspaces are degenerate and are both described by $\bar{H}_a$.  

The same algebra can be found in other geometries as well.  Consider, for example, the system in Fig.~\ref{Fig_1}b.  The Hamiltonian that describes this system is,
\begin{eqnarray}
H_b=it_{ad}\gamma_{d}\gamma_a+it_{bd}\gamma_{b}\gamma_d+it_{cd}\gamma_c\gamma_d.
\end{eqnarray}
At first glance, these operators do not obey the quaternion algebra, for example, $(i\gamma_d\gamma_a)(i\gamma_b\gamma_d)\neq\pm\gamma_c\gamma_d$.  However, since parity is conserved, the total parity operator $P=\gamma_a\gamma_b\gamma_c\gamma_d$ is simply a number (either $\pm1$).  Thus, we can replace pairs of operators with the opposite pair (e.g. $\gamma_d\gamma_a=P\gamma_b\gamma_c$).  Then the Hamiltonian can be rewritten as
\begin{eqnarray}
H_b=iP(t_{ad}\gamma_{b}\gamma_c+t_{bd}\gamma_{a}\gamma_c+t_{cd}\gamma_a\gamma_b)
\end{eqnarray}
which is simply the Hamiltonian in Eq.~\ref{Ha} and is therefore a representation of the Weyl Hamiltonian.  

Of course these are systems of separated MZMs.  In systems of quantum dots, we do not expect Majoranas to fully separate.   However, Weyl points still emerge near near "maximally separated MZMs". As we study chains of only a few quantum dots, the MZMs do not separate over a continuous range of parameters but only at discrete points.  One such point can be seen in Fig.~\ref{Fig_7}  where we claimed that the MZMs are maximally separated.  Let us now clarify what we mean by maximal separation.  We define a pair of Majorana operators as ($\gamma_{x,i\sigma},~\gamma_{y,i\sigma}$) such that $c_{i\sigma}=1/2(\gamma_{x,i\sigma}+i\gamma_{y,i\sigma})$ destroys an electron on dot $i$ with spin $\sigma$.  Furthermore, let $\ket{\psi_0}$ be the ground state of our system and $\ket{\psi_1}$ be the first excited state.  Then we can decompose the excited state into the two types of Majorana operators.  We define the two probability distributions,
\begin{eqnarray}
P_{x,i\sigma}=|\bra{\psi_1}\gamma_{x,i\sigma}\ket{\psi_0}|^2 
\nonumber \\
P_{y,i\sigma}=|\bra{\psi_1}\gamma_{y,i\sigma}\ket{\psi_0}|^2.
\end{eqnarray}
By maximal separation, we mean that we are at the point in parameter space where these two distributions overlap the least.  

\begin{figure}[t]
\begin{center}
\includegraphics[width=\columnwidth]{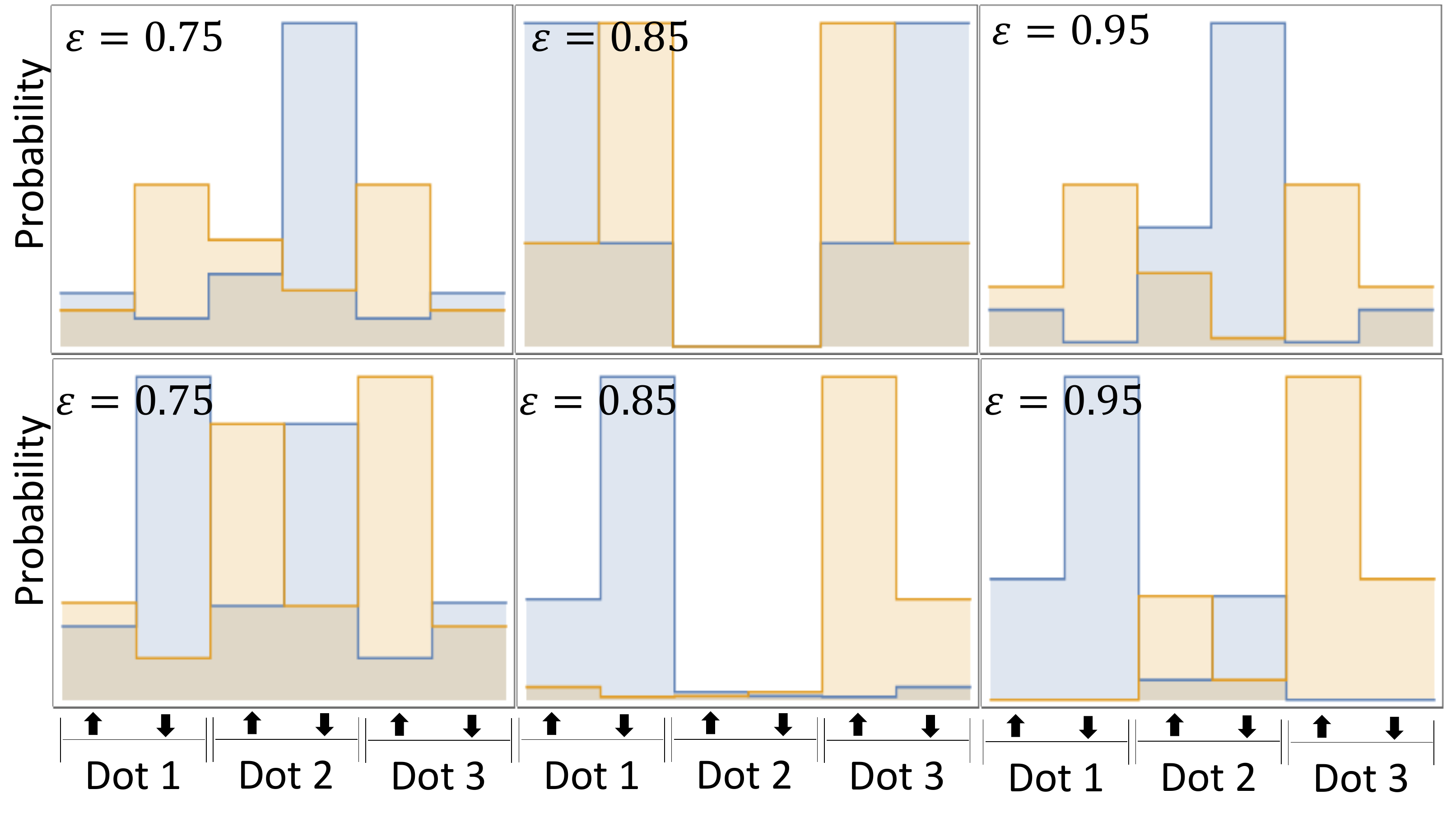}
\end{center}
\vspace{-2mm}

\caption{Probability distribution of the two Majorana decompositions.  The top row shows the probability distribution $P_{x,i\sigma}$ (yellow) and $P_{y,i\sigma}$ (blue) for a system with a magnetic field gradient as in the main text.  Here the Majoranas cannot separate to different dots yet they still separate into different spin states at $\epsilon=\epsilon_1=\epsilon_2=\epsilon_3=0.85$.  For comparison, in the bottom row we show an identical system with the exception that the magnetic field is uniform.  Here the two Majorana decompositions separate into the dots at the opposite ends of the chain.  $\epsilon$ is given in units of $2\Delta$.}    
\label{Fig_9}
\vspace{-3mm}
\end{figure}

In Fig.~\ref{Fig_9} we show the probability distributions for two different systems as a function of the potential $\epsilon_1=\epsilon_2=\epsilon_3=\epsilon$ on the dots.  The top row shows the probability distributions for the spinful 3-dot system of the main text.  Because there is a magnetic field gradient, the MZMs cannot separate to different dots, however, at $\epsilon=1.7 \Delta$ the MZMs maximally separate by occupying different spin sectors.  The bottom row shows an identical system with the exception that the magnetic field is uniform.  We see that $\epsilon=1.7 \Delta$ is indeed the point of maximal separation.  As the MZMs have more room to move in this case, they separate to either side of the chain.   

\section{Conclusion}
\label{C}

We have shown that Weyl points arise naturally in multi-quantum dot systems with superconducting leads tuned to the vicinity of maximally separated MZMs.  We propose a measurement scheme wherein current is tunneled into the first dot in a three dot chain.  The Weyl point is visualized as the crossing of $dI/dV$ peaks at a single point in the 3-dimensional parameter space.  Generically, we expect $dI/dV$ peaks to cross along 2-dimensional sheets in a 3-dimensional parameter space.   If instead the peaks close at only a single point, then the parameter space can be used to control each Pauli matrix individually and, therefore, any perturbation to the system can be corrected by the control parameters.  

We thank Sergey Frolov for helpful discussions.  This work is supported by NSF PIRE-1743717.

\bibliographystyle{apsrev4-1}
\bibliography{REF}

\begin{thebibliography}{32}%
\makeatletter
\providecommand \@ifxundefined [1]{%
 \@ifx{#1\undefined}
}%
\providecommand \@ifnum [1]{%
 \ifnum #1\expandafter \@firstoftwo
 \else \expandafter \@secondoftwo
 \fi
}%
\providecommand \@ifx [1]{%
 \ifx #1\expandafter \@firstoftwo
 \else \expandafter \@secondoftwo
 \fi
}%
\providecommand \natexlab [1]{#1}%
\providecommand \enquote  [1]{``#1''}%
\providecommand \bibnamefont  [1]{#1}%
\providecommand \bibfnamefont [1]{#1}%
\providecommand \citenamefont [1]{#1}%
\providecommand \href@noop [0]{\@secondoftwo}%
\providecommand \href [0]{\begingroup \@sanitize@url \@href}%
\providecommand \@href[1]{\@@startlink{#1}\@@href}%
\providecommand \@@href[1]{\endgroup#1\@@endlink}%
\providecommand \@sanitize@url [0]{\catcode `\\12\catcode `\$12\catcode
  `\&12\catcode `\#12\catcode `\^12\catcode `\_12\catcode `\%12\relax}%
\providecommand \@@startlink[1]{}%
\providecommand \@@endlink[0]{}%
\providecommand \url  [0]{\begingroup\@sanitize@url \@url }%
\providecommand \@url [1]{\endgroup\@href {#1}{\urlprefix }}%
\providecommand \urlprefix  [0]{URL }%
\providecommand \Eprint [0]{\href }%
\providecommand \doibase [0]{http://dx.doi.org/}%
\providecommand \selectlanguage [0]{\@gobble}%
\providecommand \bibinfo  [0]{\@secondoftwo}%
\providecommand \bibfield  [0]{\@secondoftwo}%
\providecommand \translation [1]{[#1]}%
\providecommand \BibitemOpen [0]{}%
\providecommand \bibitemStop [0]{}%
\providecommand \bibitemNoStop [0]{.\EOS\space}%
\providecommand \EOS [0]{\spacefactor3000\relax}%
\providecommand \BibitemShut  [1]{\csname bibitem#1\endcsname}%
\let\auto@bib@innerbib\@empty
\bibitem [{\citenamefont {Herring}(1937)}]{Herring1937}%
  \BibitemOpen
  \bibfield  {author} {\bibinfo {author} {\bibfnamefont {C.}~\bibnamefont
  {Herring}},\ }\href {\doibase 10.1103/PhysRev.52.365} {\bibfield  {journal}
  {\bibinfo  {journal} {Phys. Rev.}\ }\textbf {\bibinfo {volume} {52}},\
  \bibinfo {pages} {365} (\bibinfo {year} {1937})}\BibitemShut {NoStop}%
\bibitem [{\citenamefont {Yang}\ \emph {et~al.}(2011)\citenamefont {Yang},
  \citenamefont {Lu},\ and\ \citenamefont {Ran}}]{Yang2011}%
  \BibitemOpen
  \bibfield  {author} {\bibinfo {author} {\bibfnamefont {K.-Y.}\ \bibnamefont
  {Yang}}, \bibinfo {author} {\bibfnamefont {Y.-M.}\ \bibnamefont {Lu}}, \ and\
  \bibinfo {author} {\bibfnamefont {Y.}~\bibnamefont {Ran}},\ }\href {\doibase
  10.1103/PhysRevB.84.075129} {\bibfield  {journal} {\bibinfo  {journal} {Phys.
  Rev. B}\ }\textbf {\bibinfo {volume} {84}},\ \bibinfo {pages} {075129}
  (\bibinfo {year} {2011})}\BibitemShut {NoStop}%
\bibitem [{\citenamefont {Burkov}\ and\ \citenamefont
  {Balents}(2011)}]{Burkov2011}%
  \BibitemOpen
  \bibfield  {author} {\bibinfo {author} {\bibfnamefont {A.~A.}\ \bibnamefont
  {Burkov}}\ and\ \bibinfo {author} {\bibfnamefont {L.}~\bibnamefont
  {Balents}},\ }\href {\doibase 10.1103/PhysRevLett.107.127205} {\bibfield
  {journal} {\bibinfo  {journal} {Phys. Rev. Lett.}\ }\textbf {\bibinfo
  {volume} {107}},\ \bibinfo {pages} {127205} (\bibinfo {year}
  {2011})}\BibitemShut {NoStop}%
\bibitem [{\citenamefont {Xu}\ \emph {et~al.}(2015)\citenamefont {Xu},
  \citenamefont {Belopolski}, \citenamefont {Alidoust}, \citenamefont
  {Neupane}, \citenamefont {Bian}, \citenamefont {Zhang}, \citenamefont
  {Sankar}, \citenamefont {Chang}, \citenamefont {Yuan}, \citenamefont {Lee},
  \citenamefont {Huang}, \citenamefont {Zheng}, \citenamefont {Ma},
  \citenamefont {Sanchez}, \citenamefont {Wang}, \citenamefont {Bansil},
  \citenamefont {Chou}, \citenamefont {Shibayev}, \citenamefont {Lin},
  \citenamefont {Jia},\ and\ \citenamefont {Hasan}}]{Xu2015}%
  \BibitemOpen
  \bibfield  {author} {\bibinfo {author} {\bibfnamefont {S.-Y.}\ \bibnamefont
  {Xu}}, \bibinfo {author} {\bibfnamefont {I.}~\bibnamefont {Belopolski}},
  \bibinfo {author} {\bibfnamefont {N.}~\bibnamefont {Alidoust}}, \bibinfo
  {author} {\bibfnamefont {M.}~\bibnamefont {Neupane}}, \bibinfo {author}
  {\bibfnamefont {G.}~\bibnamefont {Bian}}, \bibinfo {author} {\bibfnamefont
  {C.}~\bibnamefont {Zhang}}, \bibinfo {author} {\bibfnamefont
  {R.}~\bibnamefont {Sankar}}, \bibinfo {author} {\bibfnamefont
  {G.}~\bibnamefont {Chang}}, \bibinfo {author} {\bibfnamefont
  {Z.}~\bibnamefont {Yuan}}, \bibinfo {author} {\bibfnamefont {C.-C.}\
  \bibnamefont {Lee}}, \bibinfo {author} {\bibfnamefont {S.-M.}\ \bibnamefont
  {Huang}}, \bibinfo {author} {\bibfnamefont {H.}~\bibnamefont {Zheng}},
  \bibinfo {author} {\bibfnamefont {J.}~\bibnamefont {Ma}}, \bibinfo {author}
  {\bibfnamefont {D.~S.}\ \bibnamefont {Sanchez}}, \bibinfo {author}
  {\bibfnamefont {B.}~\bibnamefont {Wang}}, \bibinfo {author} {\bibfnamefont
  {A.}~\bibnamefont {Bansil}}, \bibinfo {author} {\bibfnamefont
  {F.}~\bibnamefont {Chou}}, \bibinfo {author} {\bibfnamefont {P.~P.}\
  \bibnamefont {Shibayev}}, \bibinfo {author} {\bibfnamefont {H.}~\bibnamefont
  {Lin}}, \bibinfo {author} {\bibfnamefont {S.}~\bibnamefont {Jia}}, \ and\
  \bibinfo {author} {\bibfnamefont {M.~Z.}\ \bibnamefont {Hasan}},\ }\href
  {\doibase 10.1126/science.aaa9297} {\bibfield  {journal} {\bibinfo  {journal}
  {Science}\ }\textbf {\bibinfo {volume} {349}},\ \bibinfo {pages} {613}
  (\bibinfo {year} {2015})}\BibitemShut {NoStop}%
\bibitem [{\citenamefont {Lv}\ \emph {et~al.}(2015)\citenamefont {Lv},
  \citenamefont {Weng}, \citenamefont {Fu}, \citenamefont {Wang}, \citenamefont
  {Miao}, \citenamefont {Ma}, \citenamefont {Richard}, \citenamefont {Huang},
  \citenamefont {Zhao}, \citenamefont {Chen}, \citenamefont {Fang},
  \citenamefont {Dai}, \citenamefont {Qian},\ and\ \citenamefont
  {Ding}}]{LV2015}%
  \BibitemOpen
  \bibfield  {author} {\bibinfo {author} {\bibfnamefont {B.~Q.}\ \bibnamefont
  {Lv}}, \bibinfo {author} {\bibfnamefont {H.~M.}\ \bibnamefont {Weng}},
  \bibinfo {author} {\bibfnamefont {B.~B.}\ \bibnamefont {Fu}}, \bibinfo
  {author} {\bibfnamefont {X.~P.}\ \bibnamefont {Wang}}, \bibinfo {author}
  {\bibfnamefont {H.}~\bibnamefont {Miao}}, \bibinfo {author} {\bibfnamefont
  {J.}~\bibnamefont {Ma}}, \bibinfo {author} {\bibfnamefont {P.}~\bibnamefont
  {Richard}}, \bibinfo {author} {\bibfnamefont {X.~C.}\ \bibnamefont {Huang}},
  \bibinfo {author} {\bibfnamefont {L.~X.}\ \bibnamefont {Zhao}}, \bibinfo
  {author} {\bibfnamefont {G.~F.}\ \bibnamefont {Chen}}, \bibinfo {author}
  {\bibfnamefont {Z.}~\bibnamefont {Fang}}, \bibinfo {author} {\bibfnamefont
  {X.}~\bibnamefont {Dai}}, \bibinfo {author} {\bibfnamefont {T.}~\bibnamefont
  {Qian}}, \ and\ \bibinfo {author} {\bibfnamefont {H.}~\bibnamefont {Ding}},\
  }\href {\doibase 10.1103/PhysRevX.5.031013} {\bibfield  {journal} {\bibinfo
  {journal} {Phys. Rev. X}\ }\textbf {\bibinfo {volume} {5}},\ \bibinfo {pages}
  {031013} (\bibinfo {year} {2015})}\BibitemShut {NoStop}%
\bibitem [{\citenamefont {Lu}\ \emph {et~al.}(2015)\citenamefont {Lu},
  \citenamefont {Wang}, \citenamefont {Ye}, \citenamefont {Ran}, \citenamefont
  {Fu}, \citenamefont {Joannopoulos},\ and\ \citenamefont {Solja{\v
  c}i{\'c}}}]{Lu2015}%
  \BibitemOpen
  \bibfield  {author} {\bibinfo {author} {\bibfnamefont {L.}~\bibnamefont
  {Lu}}, \bibinfo {author} {\bibfnamefont {Z.}~\bibnamefont {Wang}}, \bibinfo
  {author} {\bibfnamefont {D.}~\bibnamefont {Ye}}, \bibinfo {author}
  {\bibfnamefont {L.}~\bibnamefont {Ran}}, \bibinfo {author} {\bibfnamefont
  {L.}~\bibnamefont {Fu}}, \bibinfo {author} {\bibfnamefont {J.~D.}\
  \bibnamefont {Joannopoulos}}, \ and\ \bibinfo {author} {\bibfnamefont
  {M.}~\bibnamefont {Solja{\v c}i{\'c}}},\ }\href {\doibase
  10.1126/science.aaa9273} {\bibfield  {journal} {\bibinfo  {journal}
  {Science}\ }\textbf {\bibinfo {volume} {349}},\ \bibinfo {pages} {622}
  (\bibinfo {year} {2015})}\BibitemShut {NoStop}%
\bibitem [{\citenamefont {Meyer}\ and\ \citenamefont
  {Houzet}(2017)}]{Meyer2017}%
  \BibitemOpen
  \bibfield  {author} {\bibinfo {author} {\bibfnamefont {J.~S.}\ \bibnamefont
  {Meyer}}\ and\ \bibinfo {author} {\bibfnamefont {M.}~\bibnamefont {Houzet}},\
  }\href {\doibase 10.1103/PhysRevLett.119.136807} {\bibfield  {journal}
  {\bibinfo  {journal} {Phys. Rev. Lett.}\ }\textbf {\bibinfo {volume} {119}},\
  \bibinfo {pages} {136807} (\bibinfo {year} {2017})}\BibitemShut {NoStop}%
\bibitem [{\citenamefont {Riwar}\ \emph {et~al.}(2016)\citenamefont {Riwar},
  \citenamefont {Houzet}, \citenamefont {Meyer},\ and\ \citenamefont
  {Nazarov}}]{Pascal2016}%
  \BibitemOpen
  \bibfield  {author} {\bibinfo {author} {\bibfnamefont {R.-P.}\ \bibnamefont
  {Riwar}}, \bibinfo {author} {\bibfnamefont {M.}~\bibnamefont {Houzet}},
  \bibinfo {author} {\bibfnamefont {J.~S.}\ \bibnamefont {Meyer}}, \ and\
  \bibinfo {author} {\bibfnamefont {Y.~V.}\ \bibnamefont {Nazarov}},\ }\href
  {\doibase 10.1038/ncomms11167} {\bibfield  {journal} {\bibinfo  {journal}
  {Nature Communictaions}\ }\textbf {\bibinfo {volume} {7}},\ \bibinfo {pages}
  {11167} (\bibinfo {year} {2016})}\BibitemShut {NoStop}%
\bibitem [{\citenamefont {Xie}\ \emph {et~al.}(2017)\citenamefont {Xie},
  \citenamefont {Vavilov},\ and\ \citenamefont {Levchenko}}]{Xie2017}%
  \BibitemOpen
  \bibfield  {author} {\bibinfo {author} {\bibfnamefont {H.-Y.}\ \bibnamefont
  {Xie}}, \bibinfo {author} {\bibfnamefont {M.~G.}\ \bibnamefont {Vavilov}}, \
  and\ \bibinfo {author} {\bibfnamefont {A.}~\bibnamefont {Levchenko}},\ }\href
  {\doibase 10.1103/PhysRevB.96.161406} {\bibfield  {journal} {\bibinfo
  {journal} {Phys. Rev. B}\ }\textbf {\bibinfo {volume} {96}},\ \bibinfo
  {pages} {161406} (\bibinfo {year} {2017})}\BibitemShut {NoStop}%
\bibitem [{\citenamefont {Xie}\ \emph {et~al.}(2018)\citenamefont {Xie},
  \citenamefont {Vavilov},\ and\ \citenamefont {Levchenko}}]{Xie2018}%
  \BibitemOpen
  \bibfield  {author} {\bibinfo {author} {\bibfnamefont {H.-Y.}\ \bibnamefont
  {Xie}}, \bibinfo {author} {\bibfnamefont {M.~G.}\ \bibnamefont {Vavilov}}, \
  and\ \bibinfo {author} {\bibfnamefont {A.}~\bibnamefont {Levchenko}},\ }\href
  {\doibase 10.1103/PhysRevB.97.035443} {\bibfield  {journal} {\bibinfo
  {journal} {Phys. Rev. B}\ }\textbf {\bibinfo {volume} {97}},\ \bibinfo
  {pages} {035443} (\bibinfo {year} {2018})}\BibitemShut {NoStop}%
\bibitem [{\citenamefont {Yokoyama}\ and\ \citenamefont
  {Nazarov}(2015)}]{Yokotama2015}%
  \BibitemOpen
  \bibfield  {author} {\bibinfo {author} {\bibfnamefont {T.}~\bibnamefont
  {Yokoyama}}\ and\ \bibinfo {author} {\bibfnamefont {Y.~V.}\ \bibnamefont
  {Nazarov}},\ }\href {\doibase 10.1103/PhysRevB.92.155437} {\bibfield
  {journal} {\bibinfo  {journal} {Phys. Rev. B}\ }\textbf {\bibinfo {volume}
  {92}},\ \bibinfo {pages} {155437} (\bibinfo {year} {2015})}\BibitemShut
  {NoStop}%
\bibitem [{\citenamefont {Scherübl}\ \emph {et~al.}(2018)\citenamefont
  {Scherübl}, \citenamefont {Pályi}, \citenamefont {Frank}, \citenamefont
  {Lukács}, \citenamefont {Fülöp}, \citenamefont {Fülöp}, \citenamefont
  {Nygård}, \citenamefont {Watanabe}, \citenamefont {Taniguchi}, \citenamefont
  {Zaránd},\ and\ \citenamefont {Csonka}}]{Zoltan2018}%
  \BibitemOpen
  \bibfield  {author} {\bibinfo {author} {\bibfnamefont {Z.}~\bibnamefont
  {Scherübl}}, \bibinfo {author} {\bibfnamefont {A.}~\bibnamefont {Pályi}},
  \bibinfo {author} {\bibfnamefont {G.}~\bibnamefont {Frank}}, \bibinfo
  {author} {\bibfnamefont {I.}~\bibnamefont {Lukács}}, \bibinfo {author}
  {\bibfnamefont {G.}~\bibnamefont {Fülöp}}, \bibinfo {author} {\bibfnamefont
  {B.}~\bibnamefont {Fülöp}}, \bibinfo {author} {\bibfnamefont
  {J.}~\bibnamefont {Nygård}}, \bibinfo {author} {\bibfnamefont
  {K.}~\bibnamefont {Watanabe}}, \bibinfo {author} {\bibfnamefont
  {T.}~\bibnamefont {Taniguchi}}, \bibinfo {author} {\bibfnamefont
  {G.}~\bibnamefont {Zaránd}}, \ and\ \bibinfo {author} {\bibfnamefont
  {S.}~\bibnamefont {Csonka}},\ }\href@noop {} {\bibfield  {journal} {\bibinfo
  {journal} {arXiv:1804.06447}\ } (\bibinfo {year} {2018})}\BibitemShut
  {NoStop}%
\bibitem [{\citenamefont {Kitaev}(2001)}]{Kitaev2001}%
  \BibitemOpen
  \bibfield  {author} {\bibinfo {author} {\bibfnamefont {A.~Y.}\ \bibnamefont
  {Kitaev}},\ }\href {\doibase 10.1070/1063-7869/44/10s/s29} {\bibfield
  {journal} {\bibinfo  {journal} {Physics-Uspekhi}\ }\textbf {\bibinfo {volume}
  {44}},\ \bibinfo {pages} {131} (\bibinfo {year} {2001})}\BibitemShut
  {NoStop}%
\bibitem [{\citenamefont {Nayak}\ \emph {et~al.}(2008)\citenamefont {Nayak},
  \citenamefont {Simon}, \citenamefont {Stern}, \citenamefont {Freedman},\ and\
  \citenamefont {Das~Sarma}}]{Nayak2008}%
  \BibitemOpen
  \bibfield  {author} {\bibinfo {author} {\bibfnamefont {C.}~\bibnamefont
  {Nayak}}, \bibinfo {author} {\bibfnamefont {S.~H.}\ \bibnamefont {Simon}},
  \bibinfo {author} {\bibfnamefont {A.}~\bibnamefont {Stern}}, \bibinfo
  {author} {\bibfnamefont {M.}~\bibnamefont {Freedman}}, \ and\ \bibinfo
  {author} {\bibfnamefont {S.}~\bibnamefont {Das~Sarma}},\ }\href {\doibase
  10.1103/RevModPhys.80.1083} {\bibfield  {journal} {\bibinfo  {journal} {Rev.
  Mod. Phys.}\ }\textbf {\bibinfo {volume} {80}},\ \bibinfo {pages} {1083}
  (\bibinfo {year} {2008})}\BibitemShut {NoStop}%
\bibitem [{\citenamefont {Beenakker}(2013)}]{Beenakker2013}%
  \BibitemOpen
  \bibfield  {author} {\bibinfo {author} {\bibfnamefont {C.}~\bibnamefont
  {Beenakker}},\ }\href {\doibase 10.1146/annurev-conmatphys-030212-184337}
  {\bibfield  {journal} {\bibinfo  {journal} {Annual Review of Condensed Matter
  Physics}\ }\textbf {\bibinfo {volume} {4}},\ \bibinfo {pages} {113} (\bibinfo
  {year} {2013})}\BibitemShut {NoStop}%
\bibitem [{\citenamefont {Alicea}(2012)}]{Alicea2012}%
  \BibitemOpen
  \bibfield  {author} {\bibinfo {author} {\bibfnamefont {J.}~\bibnamefont
  {Alicea}},\ }\href {\doibase 10.1088/0034-4885/75/7/076501} {\bibfield
  {journal} {\bibinfo  {journal} {Reports on Progress in Physics}\ }\textbf
  {\bibinfo {volume} {75}},\ \bibinfo {pages} {076501} (\bibinfo {year}
  {2012})}\BibitemShut {NoStop}%
\bibitem [{\citenamefont {Sau}\ \emph {et~al.}(2010{\natexlab{a}})\citenamefont
  {Sau}, \citenamefont {Lutchyn}, \citenamefont {Tewari},\ and\ \citenamefont
  {Das~Sarma}}]{Sau2010}%
  \BibitemOpen
  \bibfield  {author} {\bibinfo {author} {\bibfnamefont {J.~D.}\ \bibnamefont
  {Sau}}, \bibinfo {author} {\bibfnamefont {R.~M.}\ \bibnamefont {Lutchyn}},
  \bibinfo {author} {\bibfnamefont {S.}~\bibnamefont {Tewari}}, \ and\ \bibinfo
  {author} {\bibfnamefont {S.}~\bibnamefont {Das~Sarma}},\ }\href {\doibase
  10.1103/PhysRevLett.104.040502} {\bibfield  {journal} {\bibinfo  {journal}
  {Phys. Rev. Lett.}\ }\textbf {\bibinfo {volume} {104}},\ \bibinfo {pages}
  {040502} (\bibinfo {year} {2010}{\natexlab{a}})}\BibitemShut {NoStop}%
\bibitem [{\citenamefont {Alicea}(2010)}]{Alicea2010}%
  \BibitemOpen
  \bibfield  {author} {\bibinfo {author} {\bibfnamefont {J.}~\bibnamefont
  {Alicea}},\ }\href {\doibase 10.1103/PhysRevB.81.125318} {\bibfield
  {journal} {\bibinfo  {journal} {Phys. Rev. B}\ }\textbf {\bibinfo {volume}
  {81}},\ \bibinfo {pages} {125318} (\bibinfo {year} {2010})}\BibitemShut
  {NoStop}%
\bibitem [{\citenamefont {Lutchyn}\ \emph {et~al.}(2010)\citenamefont
  {Lutchyn}, \citenamefont {Sau},\ and\ \citenamefont
  {Das~Sarma}}]{Lutchyn2010}%
  \BibitemOpen
  \bibfield  {author} {\bibinfo {author} {\bibfnamefont {R.~M.}\ \bibnamefont
  {Lutchyn}}, \bibinfo {author} {\bibfnamefont {J.~D.}\ \bibnamefont {Sau}}, \
  and\ \bibinfo {author} {\bibfnamefont {S.}~\bibnamefont {Das~Sarma}},\ }\href
  {\doibase 10.1103/PhysRevLett.105.077001} {\bibfield  {journal} {\bibinfo
  {journal} {Phys. Rev. Lett.}\ }\textbf {\bibinfo {volume} {105}},\ \bibinfo
  {pages} {077001} (\bibinfo {year} {2010})}\BibitemShut {NoStop}%
\bibitem [{\citenamefont {Oreg}\ \emph {et~al.}(2010)\citenamefont {Oreg},
  \citenamefont {Refael},\ and\ \citenamefont {von Oppen}}]{Oreg2010}%
  \BibitemOpen
  \bibfield  {author} {\bibinfo {author} {\bibfnamefont {Y.}~\bibnamefont
  {Oreg}}, \bibinfo {author} {\bibfnamefont {G.}~\bibnamefont {Refael}}, \ and\
  \bibinfo {author} {\bibfnamefont {F.}~\bibnamefont {von Oppen}},\ }\href
  {\doibase 10.1103/PhysRevLett.105.177002} {\bibfield  {journal} {\bibinfo
  {journal} {Phys. Rev. Lett.}\ }\textbf {\bibinfo {volume} {105}},\ \bibinfo
  {pages} {177002} (\bibinfo {year} {2010})}\BibitemShut {NoStop}%
\bibitem [{\citenamefont {Sau}\ \emph {et~al.}(2010{\natexlab{b}})\citenamefont
  {Sau}, \citenamefont {Tewari}, \citenamefont {Lutchyn}, \citenamefont
  {Stanescu},\ and\ \citenamefont {Das~Sarma}}]{Sau2010b}%
  \BibitemOpen
  \bibfield  {author} {\bibinfo {author} {\bibfnamefont {J.~D.}\ \bibnamefont
  {Sau}}, \bibinfo {author} {\bibfnamefont {S.}~\bibnamefont {Tewari}},
  \bibinfo {author} {\bibfnamefont {R.~M.}\ \bibnamefont {Lutchyn}}, \bibinfo
  {author} {\bibfnamefont {T.~D.}\ \bibnamefont {Stanescu}}, \ and\ \bibinfo
  {author} {\bibfnamefont {S.}~\bibnamefont {Das~Sarma}},\ }\href {\doibase
  10.1103/PhysRevB.82.214509} {\bibfield  {journal} {\bibinfo  {journal} {Phys.
  Rev. B}\ }\textbf {\bibinfo {volume} {82}},\ \bibinfo {pages} {214509}
  (\bibinfo {year} {2010}{\natexlab{b}})}\BibitemShut {NoStop}%
\bibitem [{\citenamefont {Ptok}\ \emph {et~al.}(2017)\citenamefont {Ptok},
  \citenamefont {Kobia\l{}ka},\ and\ \citenamefont {Doma\ifmmode~\acute{n}\else
  \'{n}\fi{}ski}}]{Ptok2017}%
  \BibitemOpen
  \bibfield  {author} {\bibinfo {author} {\bibfnamefont {A.}~\bibnamefont
  {Ptok}}, \bibinfo {author} {\bibfnamefont {A.}~\bibnamefont {Kobia\l{}ka}}, \
  and\ \bibinfo {author} {\bibfnamefont {T.}~\bibnamefont
  {Doma\ifmmode~\acute{n}\else \'{n}\fi{}ski}},\ }\href {\doibase
  10.1103/PhysRevB.96.195430} {\bibfield  {journal} {\bibinfo  {journal} {Phys.
  Rev. B}\ }\textbf {\bibinfo {volume} {96}},\ \bibinfo {pages} {195430}
  (\bibinfo {year} {2017})}\BibitemShut {NoStop}%
\bibitem [{\citenamefont {Mourik}\ \emph {et~al.}(2012)\citenamefont {Mourik},
  \citenamefont {Zuo}, \citenamefont {Frolov}, \citenamefont {Plissard},
  \citenamefont {Bakkers},\ and\ \citenamefont {Kouwenhoven}}]{Mourik2012}%
  \BibitemOpen
  \bibfield  {author} {\bibinfo {author} {\bibfnamefont {V.}~\bibnamefont
  {Mourik}}, \bibinfo {author} {\bibfnamefont {K.}~\bibnamefont {Zuo}},
  \bibinfo {author} {\bibfnamefont {S.~M.}\ \bibnamefont {Frolov}}, \bibinfo
  {author} {\bibfnamefont {S.~R.}\ \bibnamefont {Plissard}}, \bibinfo {author}
  {\bibfnamefont {E.~P. A.~M.}\ \bibnamefont {Bakkers}}, \ and\ \bibinfo
  {author} {\bibfnamefont {L.~P.}\ \bibnamefont {Kouwenhoven}},\ }\href
  {\doibase 10.1126/science.1222360} {\bibfield  {journal} {\bibinfo  {journal}
  {Science}\ }\textbf {\bibinfo {volume} {336}},\ \bibinfo {pages} {1003}
  (\bibinfo {year} {2012})}\BibitemShut {NoStop}%
\bibitem [{\citenamefont {Chen}\ \emph {et~al.}(2017)\citenamefont {Chen},
  \citenamefont {Yu}, \citenamefont {Stenger}, \citenamefont {Hocevar},
  \citenamefont {Car}, \citenamefont {Plissard}, \citenamefont {Bakkers},
  \citenamefont {Stanescu},\ and\ \citenamefont {Frolov}}]{Chen2017}%
  \BibitemOpen
  \bibfield  {author} {\bibinfo {author} {\bibfnamefont {J.}~\bibnamefont
  {Chen}}, \bibinfo {author} {\bibfnamefont {P.}~\bibnamefont {Yu}}, \bibinfo
  {author} {\bibfnamefont {J.}~\bibnamefont {Stenger}}, \bibinfo {author}
  {\bibfnamefont {M.}~\bibnamefont {Hocevar}}, \bibinfo {author} {\bibfnamefont
  {D.}~\bibnamefont {Car}}, \bibinfo {author} {\bibfnamefont {S.~R.}\
  \bibnamefont {Plissard}}, \bibinfo {author} {\bibfnamefont {E.~P. A.~M.}\
  \bibnamefont {Bakkers}}, \bibinfo {author} {\bibfnamefont {T.~D.}\
  \bibnamefont {Stanescu}}, \ and\ \bibinfo {author} {\bibfnamefont {S.~M.}\
  \bibnamefont {Frolov}},\ }\href {\doibase 10.1126/sciadv.1701476} {\bibfield
  {journal} {\bibinfo  {journal} {Science Advances}\ }\textbf {\bibinfo
  {volume} {3}} (\bibinfo {year} {2017}),\ 10.1126/sciadv.1701476}\BibitemShut
  {NoStop}%
\bibitem [{\citenamefont {Deng}\ \emph {et~al.}(2016)\citenamefont {Deng},
  \citenamefont {Vaitiekenas}, \citenamefont {Hansen}, \citenamefont {Danon},
  \citenamefont {Leijnse}, \citenamefont {Flensberg}, \citenamefont {Nyg{\r
  a}rd}, \citenamefont {Krogstrup},\ and\ \citenamefont {Marcus}}]{Deng2016}%
  \BibitemOpen
  \bibfield  {author} {\bibinfo {author} {\bibfnamefont {M.~T.}\ \bibnamefont
  {Deng}}, \bibinfo {author} {\bibfnamefont {S.}~\bibnamefont {Vaitiekenas}},
  \bibinfo {author} {\bibfnamefont {E.~B.}\ \bibnamefont {Hansen}}, \bibinfo
  {author} {\bibfnamefont {J.}~\bibnamefont {Danon}}, \bibinfo {author}
  {\bibfnamefont {M.}~\bibnamefont {Leijnse}}, \bibinfo {author} {\bibfnamefont
  {K.}~\bibnamefont {Flensberg}}, \bibinfo {author} {\bibfnamefont
  {J.}~\bibnamefont {Nyg{\r a}rd}}, \bibinfo {author} {\bibfnamefont
  {P.}~\bibnamefont {Krogstrup}}, \ and\ \bibinfo {author} {\bibfnamefont
  {C.~M.}\ \bibnamefont {Marcus}},\ }\href {\doibase 10.1126/science.aaf3961}
  {\bibfield  {journal} {\bibinfo  {journal} {Science}\ }\textbf {\bibinfo
  {volume} {354}},\ \bibinfo {pages} {1557} (\bibinfo {year}
  {2016})}\BibitemShut {NoStop}%
\bibitem [{\citenamefont {Nadj-Perge}\ \emph {et~al.}(2014)\citenamefont
  {Nadj-Perge}, \citenamefont {Drozdov}, \citenamefont {Li}, \citenamefont
  {Chen}, \citenamefont {Jeon}, \citenamefont {Seo}, \citenamefont {MacDonald},
  \citenamefont {Bernevig},\ and\ \citenamefont {Yazdani}}]{Nadj2014}%
  \BibitemOpen
  \bibfield  {author} {\bibinfo {author} {\bibfnamefont {S.}~\bibnamefont
  {Nadj-Perge}}, \bibinfo {author} {\bibfnamefont {I.~K.}\ \bibnamefont
  {Drozdov}}, \bibinfo {author} {\bibfnamefont {J.}~\bibnamefont {Li}},
  \bibinfo {author} {\bibfnamefont {H.}~\bibnamefont {Chen}}, \bibinfo {author}
  {\bibfnamefont {S.}~\bibnamefont {Jeon}}, \bibinfo {author} {\bibfnamefont
  {J.}~\bibnamefont {Seo}}, \bibinfo {author} {\bibfnamefont {A.~H.}\
  \bibnamefont {MacDonald}}, \bibinfo {author} {\bibfnamefont {B.~A.}\
  \bibnamefont {Bernevig}}, \ and\ \bibinfo {author} {\bibfnamefont
  {A.}~\bibnamefont {Yazdani}},\ }\href {\doibase 10.1126/science.1259327}
  {\bibfield  {journal} {\bibinfo  {journal} {Science}\ }\textbf {\bibinfo
  {volume} {346}},\ \bibinfo {pages} {602} (\bibinfo {year}
  {2014})}\BibitemShut {NoStop}%
\bibitem [{\citenamefont {Fulga}\ \emph {et~al.}(2013)\citenamefont {Fulga},
  \citenamefont {Haim}, \citenamefont {Akhmerov},\ and\ \citenamefont
  {Oreg}}]{Fulga2013}%
  \BibitemOpen
  \bibfield  {author} {\bibinfo {author} {\bibfnamefont {I.~C.}\ \bibnamefont
  {Fulga}}, \bibinfo {author} {\bibfnamefont {A.}~\bibnamefont {Haim}},
  \bibinfo {author} {\bibfnamefont {A.~R.}\ \bibnamefont {Akhmerov}}, \ and\
  \bibinfo {author} {\bibfnamefont {Y.}~\bibnamefont {Oreg}},\ }\href {\doibase
  10.1088/1367-2630/15/4/045020} {\bibfield  {journal} {\bibinfo  {journal}
  {New Journal of Physics}\ }\textbf {\bibinfo {volume} {15}},\ \bibinfo
  {pages} {045020} (\bibinfo {year} {2013})}\BibitemShut {NoStop}%
\bibitem [{\citenamefont {Stenger}\ \emph {et~al.}(2018)\citenamefont
  {Stenger}, \citenamefont {Woods}, \citenamefont {Frolov},\ and\ \citenamefont
  {Stanescu}}]{Stenger2018}%
  \BibitemOpen
  \bibfield  {author} {\bibinfo {author} {\bibfnamefont {J.~P.~T.}\
  \bibnamefont {Stenger}}, \bibinfo {author} {\bibfnamefont {B.~D.}\
  \bibnamefont {Woods}}, \bibinfo {author} {\bibfnamefont {S.~M.}\ \bibnamefont
  {Frolov}}, \ and\ \bibinfo {author} {\bibfnamefont {T.~D.}\ \bibnamefont
  {Stanescu}},\ }\href {\doibase 10.1103/PhysRevB.98.085407} {\bibfield
  {journal} {\bibinfo  {journal} {Phys. Rev. B}\ }\textbf {\bibinfo {volume}
  {98}},\ \bibinfo {pages} {085407} (\bibinfo {year} {2018})}\BibitemShut
  {NoStop}%
\bibitem [{\citenamefont {Zhang}\ and\ \citenamefont {Nori}(2016)}]{Zhang2016}%
  \BibitemOpen
  \bibfield  {author} {\bibinfo {author} {\bibfnamefont {P.}~\bibnamefont
  {Zhang}}\ and\ \bibinfo {author} {\bibfnamefont {F.}~\bibnamefont {Nori}},\
  }\href {\doibase 10.1088/1367-2630/18/4/043033} {\bibfield  {journal}
  {\bibinfo  {journal} {New Journal of Physics}\ }\textbf {\bibinfo {volume}
  {18}},\ \bibinfo {pages} {043033} (\bibinfo {year} {2016})}\BibitemShut
  {NoStop}%
\bibitem [{\citenamefont {Su}\ \emph {et~al.}(2017)\citenamefont {Su},
  \citenamefont {Tacla}, \citenamefont {Hocevar}, \citenamefont {Car},
  \citenamefont {Plissard}, \citenamefont {Bakkers}, \citenamefont {Daley},
  \citenamefont {Pekker},\ and\ \citenamefont {Frolov}}]{Su2017}%
  \BibitemOpen
  \bibfield  {author} {\bibinfo {author} {\bibfnamefont {Z.}~\bibnamefont
  {Su}}, \bibinfo {author} {\bibfnamefont {A.~B.}\ \bibnamefont {Tacla}},
  \bibinfo {author} {\bibfnamefont {M.}~\bibnamefont {Hocevar}}, \bibinfo
  {author} {\bibfnamefont {D.}~\bibnamefont {Car}}, \bibinfo {author}
  {\bibfnamefont {S.~R.}\ \bibnamefont {Plissard}}, \bibinfo {author}
  {\bibfnamefont {E.~P. A.~M.}\ \bibnamefont {Bakkers}}, \bibinfo {author}
  {\bibfnamefont {A.~J.}\ \bibnamefont {Daley}}, \bibinfo {author}
  {\bibfnamefont {D.}~\bibnamefont {Pekker}}, \ and\ \bibinfo {author}
  {\bibfnamefont {S.~M.}\ \bibnamefont {Frolov}},\ }\href {\doibase
  10.1038/s41467-017-00665-7} {\bibfield  {journal} {\bibinfo  {journal}
  {Nature Communications}\ }\textbf {\bibinfo {volume} {8}},\ \bibinfo {pages}
  {585} (\bibinfo {year} {2017})}\BibitemShut {NoStop}%
\bibitem [{\citenamefont {Maurer}\ \emph {et~al.}(2018)\citenamefont {Maurer},
  \citenamefont {Gamble}, \citenamefont {Tracy}, \citenamefont {Eley},\ and\
  \citenamefont {Lu}}]{Maurer2018}%
  \BibitemOpen
  \bibfield  {author} {\bibinfo {author} {\bibfnamefont {L.}~\bibnamefont
  {Maurer}}, \bibinfo {author} {\bibfnamefont {J.}~\bibnamefont {Gamble}},
  \bibinfo {author} {\bibfnamefont {L.}~\bibnamefont {Tracy}}, \bibinfo
  {author} {\bibfnamefont {S.}~\bibnamefont {Eley}}, \ and\ \bibinfo {author}
  {\bibfnamefont {T.}~\bibnamefont {Lu}},\ }\href {\doibase
  10.1103/PhysRevApplied.10.054071} {\bibfield  {journal} {\bibinfo  {journal}
  {Phys. Rev. Applied}\ }\textbf {\bibinfo {volume} {10}},\ \bibinfo {pages}
  {054071} (\bibinfo {year} {2018})}\BibitemShut {NoStop}%
\bibitem [{\citenamefont {Drndić}\ \emph {et~al.}(1998)\citenamefont
  {Drndić}, \citenamefont {Johnson}, \citenamefont {Thywissen}, \citenamefont
  {Prentiss},\ and\ \citenamefont {Westervelt}}]{Drndic1998}%
  \BibitemOpen
  \bibfield  {author} {\bibinfo {author} {\bibfnamefont {M.}~\bibnamefont
  {Drndić}}, \bibinfo {author} {\bibfnamefont {K.~S.}\ \bibnamefont
  {Johnson}}, \bibinfo {author} {\bibfnamefont {J.~H.}\ \bibnamefont
  {Thywissen}}, \bibinfo {author} {\bibfnamefont {M.}~\bibnamefont {Prentiss}},
  \ and\ \bibinfo {author} {\bibfnamefont {R.~M.}\ \bibnamefont {Westervelt}},\
  }\href {\doibase 10.1063/1.121455} {\bibfield  {journal} {\bibinfo  {journal}
  {Applied Physics Letters}\ }\textbf {\bibinfo {volume} {72}},\ \bibinfo
  {pages} {2906} (\bibinfo {year} {1998})}\BibitemShut {NoStop}%
\end{thebibliography}%

\end{document}